\documentclass[%
 reprint,
 amsmath,amssymb,
 aps,
]{revtex4-2}

\usepackage[dvipsnames]{xcolor}
\usepackage{graphicx}
\usepackage{dcolumn}
\usepackage{bm}
\usepackage{bbold}
\usepackage{amsmath,amssymb}

\begin{document}

\preprint{APS/123-QED}

\title{Time-dependent transport in Graphene Mach-Zender Interferometers}

\author{Gaia Forghieri$^1$}
\author{Paolo Bordone$^{1,2}$}%
\author{Andrea Bertoni$^2$}
\affiliation{%
 $^1$Dipartimento di Scienze Fisiche, Informatiche e Matematiche,
Università degli Studi di Modena e Reggio Emilia, Via Campi 213/A, 41125 Modena, Italy\\
 $^2$S3, Istituto Nanoscienze-CNR, Via Campi 213/A, 41125 Modena, Italy
}%





\begin{abstract}
Graphene nanoribbons provide an ideal platform for electronic interferometry in the Integer Quantum Hall regime. Here, we solve the time-dependent four-component Schr\"odinger equation for single carriers in graphene and expose several dynamical effects of the carrier localization on their transport characteristics in pn junctions. We simulate two kinds of Mach-Zender Interferometers (MZI). The first is based on Quantum Point Contacts and is similar to traditional GaAs/AlGaAs interferometers. As expected, we observe Aharonov-Bohm oscillations and phase averaging. The second is based on Valley Beam Splitters, where we observe unexpected phenomena due to the intersection of the Edge Channels that constitute the MZI. Our results provide further insights into the behavior of graphene interferometers. Additionally, they highlight the operative regime of such nanodevices for feasible single-particle implementations.\newline
\end{abstract}

\maketitle


\section{INTRODUCTION}
\label{sec:intro}

Monolayer graphene is widely known to present anomalous Half-Integer Quantum Hall Effect\cite{novoselov2005,zhang2005,ho2008,wu2009,fujita2016}, with conductance $\sigma=4\left(n+\frac{1}{2}\right)\frac{e^2}{\hbar}$. This happens because Landau Levels (LL) in the Integer Quantum Hall (IQH) regime in graphene possess SU(4) symmetry: they are fourfold degenerate in the spin and valley degree of freedom, and the zeroth LL falls on the Fermi level. When degeneracy is lifted and the SU(4) symmetry is broken, it is also possible to observe the usual Integer Quantum Hall Effect (IQHE)\cite{jiang2007,gusynin2008,lado2013,amet2014,zimmermann2017,ding2018,liu2022} with the conductance $\sigma=n\frac{e^2}{\hbar}$. LL splitting also allows to observe many kinds of collective excitations, such as magnons\cite{wei2018,assouline2021,pierce2021}, valley skyrmions\cite{lian2017}, and even strain-induced pseudomagnetoexcitons\cite{berman2022}. For these reasons, graphene has recently raised much attention in the field of electronic quantum optics\cite{wei2017,araujo2020,carrega2021,jo2021,deprez2021,johnson2021, marconcini2022, mirzakhani2022, edlbauer2022} in the Integer and Fractional Quantum Hall regimes\cite{carrega2021}, and offers promising perspectives for quantum computation as well\cite{furtado2022}.\newline 
The degeneracy lifting of LLs implies the presence of a gap between the conduction and valence band of graphene, which is in general present in experimental conditions, although specific production techniques such as encapsulation with hexagonal Boron Nitride (hBN)\cite{dean2010,wang2017} are able to produce high-purity graphene samples with essentially no gap. Very clean pn junctions in nanoribbons of monolayer graphene are then created, to study transport phenomena in the Quantum Hall regime: among these, the coexistence between Edge States and Snake States along pn junctions\cite{oroszlany2008,williams2011,carmier2011,milovanovic2014,taychatanapat2015,rickhaus2015_2,chen2016,kolasinski2017,makk2018,barbier2012}, and the occurrence of Klein tunneling through electrostatic barriers\cite{barbier2012,cheianov2006,young2009,tudorovskiy2012,araujo2020}.\newline

The gap in graphene encapsulated with e.g. hBN is very small (of the order of tens of $meV$), in contrast with the semiconductor gap in GaAs/AlGaAs heterostructures\cite{wang2017}. Thus, in graphene monolayers both electronic- and hole-like states are available for transport at the same time, even considering very small voltage gradients in the system. This has proved to be detrimental for Mach-Zender Interferometers (MZI) which exploit external metallic gates to create Quantum Point Contacts (QPC), because of back-scattering at the edge of the ribbon and/or below the electrostatic gates\cite{abanin2007,xiang2016,marguerite2019,moreau2021,kumar2022}. Instead, the most recent implementations of MZIs in graphene exploit the valley degree of freedom of the electronic wave function. This means creating devices that have no equivalent in GaAs/AlGaAs heterostructures. Because of this, graphene has become of great interest in the context of valleytronics\cite{schaibley2016}, a kind of electronics that stores information in the valley degree of freedom. Recent implementations of MZIs in monolayer graphene are in fact characterized by Valley Beam Splitters (VBS) at the edge of armchair nanoribbons\cite{morikawa2015,morikawa2015,trifunovic2019,wei2017,jo2021,johnson2021,flor2022}. The arms of these MZIs are the two co-propagating Edge Channels (one for each valley) that are formed along a pn junction when valley degeneracy is lifted for the zeroth Landau Level (LL). These kind of interferometers present visibilities up to 98\% when working with delocalized currents\cite{wei2017,jo2021}. Hybrid monolayer-bilayer interfaces also present valley polarized channels and operate in a similar manner\cite{mirzakhani2022}.\newline

In our study we simulate the dynamics of localized carriers in armchair nanoribbons of monolayer graphene. In this way, we aim to characterize the functional regimes of MZIs for future implementation in single- and few-carrier interferometry\cite{kotilahti2021, edlbauer2022}. Our approach consists in numerically integrating the exact time-dependent Schr\"odinger equation for a single electron in the IQH regime in graphene through the Split-Step Fourier Method\cite{kramer2010,chaves2015,grasselli2015,grasselli2016}. We simulate the full two-dimensional system in real space, in order to capture the behavior of the wave packet in the most realistic and immediate way. Our group has previously performed various studies for MZIs in GaAs/AlGaAs heterostructures, where the dynamics of the propagating carriers is limited to the conduction band\cite{beggi2015,beggi2015_2,bellentani2018,bordone2019,bellentani2020}. Here we include the whole band structure in the proximity of the valleys in the Brillouin Zone, and consider the electronic wave function as a four-component spinor (see Appendix \ref{sec:split-step} for a detailed derivation). This allows us to take into account the sublattice and valley degrees of freedom.

This article is organized as follows. In Sec. \ref{sec:model} we first introduce the theoretical description of the IQHE in graphene armchair nanoribbons, then we show the numerical model at the base of our simulations. In Sec. \ref{sec:snakestates} we observe the possible transport regimes of localized carriers along pn and npn junctions, and then in Sec. \ref{sec:QPC} we characterize the behavior of Quantum Point Contacts in graphene. In Sec. \ref{sec:MZI1} we describe a first MZI with QPCs and report Aharonov-Bohm oscillations and phase averaging. In Sec. \ref{sec:MZI2} we study a MZI made with VBSs and characterize its behavior in our regime for single-particle transport. Finally, in Sec. \ref{sec:conclusions} we draw our conclusions.\newline

\section{PHYSICAL SYSTEM AND NUMERICAL MODEL}
\label{sec:model}
Monolayer graphene consists of a triangular lattice with lattice vectors $\textbf{a}=a(1,0)$ and $\textbf{b}=a(1/2,\sqrt{3}/2)$. The unit cell presents two carbon atoms: the first located at $(0,0)$ belonging to sublattice $A$, and the second at $\textbf{d}=a(0,1/\sqrt{3})$ belonging to sublattice $B$. The lattice parameter is $a=\sqrt{3}\,A$, with $A=1.422\,\dot{A}$ being the Carbon-Carbon bond length\cite{hancock2010}. The last valence band and first conduction band of graphene can be found through the tight-binding model for nearest-neighbor hopping\cite{katsnelson2007,hancock2010,lado2013}. The bands touch at the corners of the hexagonal Brillouin Zone, at which points their dispersion is linear and Dirac cones are observed. Only two valleys in the BZ are inequivalent: we call them $K=\frac{2\pi}{a\sqrt{3}}\left(\frac{1}{\sqrt{3}},1\right)$ and $K'=\frac{2\pi}{a\sqrt{3}}\left(-\frac{1}{\sqrt{3}},1\right)$. In total the electronic wave function $\Psi$ is written in terms of the sublattice ($A/B$) and valley ($K/K'$) degrees of freedom. Thus, $\Psi$ is a spinor with a total of four components, $\Psi=(\psi_A^K,\psi_B^K,\psi_A^{K'},\psi_B^{K'})$.\newline
Our calculations are performed within the $\textbf{k}\cdot\textbf{p}$ approximation, in which the bands are approximated by a linear dispersion near the valleys. For the sake of simplicity of notation, in place of the absolute wave vector $\textbf{k}$ we define the relative vector with respect to $K/K'$ as $\textbf{q}=\textbf{k}-K(K')$. This way, the electronic $4\times 4$ Hamiltonian has the shape of a Dirac Hamiltonian for massless particles:
\begin{align}\label{eq:diracK}
\hat{H}&=\hbar v_F\begin{pmatrix}0 & -q_x+iq_y & 0 & 0 \\ -q_x-iq_y & 0 & 0 & 0 \\ 0 & 0 & 0 & q_x+iq_y \\ 0 & 0 & q_x -iq_y & 0\end{pmatrix}= \nonumber \\
&=\hbar v_F\left(-q_x\sigma^{AB}_x\otimes \sigma^{KK'}_z-q_y\sigma^{AB}_y\otimes \mathbb{1}^{KK'}\right),
\end{align}
where $v_F=\frac{3at}{2}\cdot \frac{1}{\hbar}$ is the Fermi velocity, with $t=2.7\, eV$ the hopping parameter of graphene\cite{hancock2010}; $\sigma^{AB}_{x/y}$ are the Pauli matrices on the sublattice isospin basis; $\sigma^{KK'}_{z}$ and $\mathbb{1}^{KK'}$ are the $z$ Pauli matrix and identity matrix on the valley isospin basis, respectively. A term equal to $M\sigma^{AB}_z\otimes\mathbb{1}^{KK'}$ can also be added, thus introducing an energy difference between sublattices A and B. This corresponds to the action of a field that couples to the sublattice degree of freedom, an important phenomenon to consider in experimental realizations of monolayer graphene samples\cite{goerbig2011,lado2013,lado2015,wang2017}. We will call $M$ a mass term, since it opens a gap of $E_g=2M$ between the valence and conduction bands and introduces an effective mass.

Note that in our treatment we neglect the real spin degree of freedom. However, it has been shown that interchannel scattering is suppressed for opposite spins\cite{amet2014}. Thus, states with opposite spins behave independently of each other and do not interfere. Consequently, the results of our simulations are valid separately for any value of the spin.

\subsection{IQH regime in Armchair Nanoribbons}
\label{sec:armchair}
A graphene ribbon is defined by an infinite length in the $y$ direction and a limited width $L$ in $x$. The edges of the ribbon are made up of unsaturated carbon atoms. In the case of an armchair nanoribbon, each edge consists of alternating atoms from both sublattices: half belong to $A$ and half to $B$. The total $A/B$ components of $\Psi$ in an armchair nanoribbon are a specific combination\cite{marconcini2010} of $\psi^{K/K'}_{A/B}$, more specifically:
\begin{equation}\label{eq:armchaircomb}
\begin{cases}
\psi_A(\textbf{r})=\frac{1}{\sqrt{2}}\left( e^{i\textbf{K}\cdot \textbf{r}}\psi_A^K(\textbf{r})-ie^{i\textbf{K}'\cdot \textbf{r}}\psi_A^{K'}(\textbf{r})\right) \\
\psi_B(\textbf{r})=\frac{1}{\sqrt{2}}\left(ie^{i\textbf{K}\cdot \textbf{r}}\psi_B^K(\textbf{r})+e^{i\textbf{K}'\cdot \textbf{r}}\psi_B^{K'}(\textbf{r}) \right)
\end{cases} .
\end{equation}
The finite range $x\in[0,L]$ of the ribbon imposes $\psi_{A/B}$ to vanish at the edges of the ribbon. Consequently, the following boundary conditions apply:
\begin{equation}\label{eq:boundary2}
\begin{cases}
\psi_A^K(0,y)=i\psi_A^{K'}(0,y) \\
\psi_B^K(0,y)=i\psi_B^{K'}(0,y) \\
\psi_A^K(L,y)=ie^{i(K'_x-K_x)L}\psi_A^{K'}(L,y) \\
\psi_B^K(L,y)=ie^{i(K'_x-K_x)L}\psi_B^{K'}(L,y)
\end{cases},
\end{equation}
where we used $K_y=K_y'$ to simplify the phase factor in $y$. The problem simplifies if we define a new spinor $\varphi$ in an extended real space $x\in[-L,L]$\cite{marconcini2010}:
\begin{equation}\label{eq:varphi}
\varphi(x,y)=\begin{cases} e^{iK_xx}\Psi^K(x,y) &0\leq x\leq L \\ e^{iK_x'x}\Psi^{K'}(-x,y) &-L\leq x<0 \end{cases}.
\end{equation}
The boundary conditions in Eq. (\ref{eq:boundary2}) directly translate to $\varphi(x,y)$ being continuous in $x=0$ and periodic in $[-L,L]$. The Hamiltonian for $\varphi$ reads:
\begin{equation}\label{eq:dirac3}
\hat{H}=\begin{pmatrix}
V(\left|x\right|,y)+M & \hbar v_F\left(-k_x+ik_y\right) \\ \hbar v_F\left(-k_x-ik_y\right) & V(\left|x\right|,y)-M
\end{pmatrix}.
\end{equation}
In the most general case $M\neq 0$ and $V(x,y)$ is an external electrostatic potential. Note that Eq. (\ref{eq:dirac3}) is written in terms of the absolute wave vector $\textbf{k}$, to take into account the relative phase factors for different valleys.\newline
We can introduce an external magnetic field $\textbf{B}=B\hat{\textbf{z}}$ by using the Landau gauge $\textbf{A}=B|x|\hat{\textbf{y}}$; the absolute value comes from the mirrored nature of our system in the $x$ direction. Then, we perform the Peierls substitution, $\textbf{k}\rightarrow \textbf{k}+e\textbf{A}/\hbar$, and consider the \textit{ansatz} $\varphi_{n,k}(x,y) = e^{iky}\varphi_{n,k}(x)$. Let's consider a translationally invariant potential, $V(x,y)=V_x(x)$ and rewrite the wave vectors as $k_{x/y}=-i\frac{\partial}{\partial x/y}$: in this way, the x-component $\varphi_{n,k}(x)$ becomes the eigenvector of an effective 1D Hamiltonian,
\begin{equation}\label{eq:dirac4}
\resizebox{.9\hsize}{!}{$
\hat{H}^{eff}=\begin{pmatrix}
V_x(\left|x\right|)+M & iv_F\left(\frac{\partial}{\partial x}+\frac{1}{l_m^2}(\left|x\right|-x_0)\right) \\ iv_F\left(\frac{\partial}{\partial x}-\frac{1}{l_m^2}(\left|x\right|-x_0)\right) & V_x(\left|x\right|)-M
\end{pmatrix}
$}.
\end{equation}
In Eq. (\ref{eq:dirac4}), $l_m=\sqrt{\hbar/eB}$ is the magnetic length; $x_0=-\hbar k/eB$ is the center of the effective linear confinement term induced by the magnetic field on the $x$ direction. Notice that $x_0$ is coupled to the quantum number $k$, which is the wave vector in the $y$ direction. The cyclotron frequency of the system is defined as $\omega_c=\sqrt{2} v_F/l_m$.\newline

The energies of the LLs in `bulk' are the eigenvalues of $\hat{H}^{eff}$ when $x_0$ is sufficiently away from the edges. If $V_x$ is constant and null\cite{goerbig2011,lado2013}:
\begin{equation}\label{eq:LL2}
E_n=\pm\sqrt{(\hbar\omega_c)^2n+M^2},
\end{equation}
and the components of the eigenstates are solutions of the quantum harmonic oscillator with frequency $\omega_c$\cite{brey2006}. Specifically, $\varphi_{n,k}(x)=(\phi_{n-1}(x-x_0),\phi_n(x-x_0))^T$ for the $K$ valley ($x_0>0$) and $\varphi_{n,k}(x)=(\phi_n(x-x_0),\phi_{n-1}(x-x_0))^T$ for the $K'$ valley ($x_0<0$), where $T$ indicates the transpose.\newline
The expression in Eq. (\ref{eq:LL2}) means that the distance between subsequent LLs decreases for higher values of $n$. This is in contrast to the traditional case in GaAs/AlGaAs heterostructures, in which $\Delta E$ remains constant\cite{bordone2019}. Additionally, there is an infinite number of negative energies for hole-like states, and for $M=0$ there is no zero-point energy and $E_0=0$. Notice also that all LLs are two times degenerate in the valley degree of freedom except for $LL_0^+$ and $LL_0^-$, which split when $M\neq 0$ to generate the expected gap of $E_g=2M$\cite{lado2013}.\newline

\begin{figure}[t]
\centering
\includegraphics[width=0.5\textwidth]{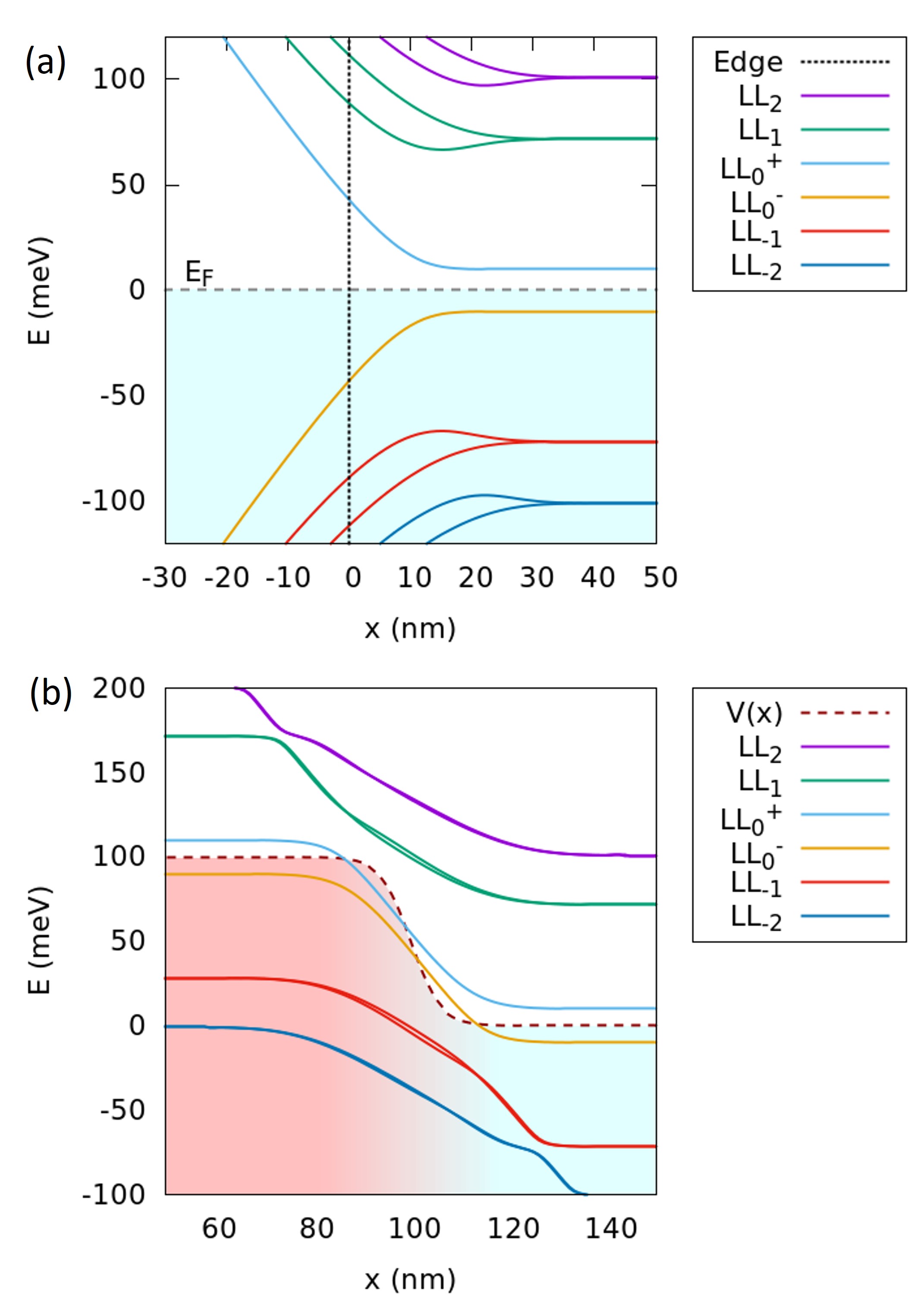}
\caption{LL of a graphene armchair nanoribbon defined in $x\in[0,200]\,nm$, with magnetic field $B=5\, T$ and mass gap $E_g=20\,meV$. (a) LL at the edge of the nanoribbon with no additional electrostatic potential. The blue region corresponds to the energies below the Fermi level. The black dotted line represents the position of the edge. (b) LL at the center of the nanoribbon in presence of a smooth electrostatic barrier (see Eq. (\ref{eq:fermi})) of height $\Delta V=100\,meV$ and slope $S=0.33\,nm^{-1}$. The red and blue regions indicate the $p$ and $n$ region.}
\label{fig:img0}
\end{figure}

\subsection{Localized Gaussian wave packets of Edge States}
\label{sec:gaussian}
When $x_0$ approaches the ribbon edges, an additional term $\epsilon(k)$ is added to Eq. (\ref{eq:LL2}). As shown in Fig.~\ref{fig:img0}(a), this causes positive LLs to bend upwards and negative LLs to bend downwards, thus generating dispersion of opposite sign. The term $\epsilon(k)$ is also known to lift the valley degeneracy of LLs at the edges of armchair nanoribbons\cite{brey2006}, as is our case. Figure \ref{fig:img0}(b) shows that a non-constant $V_x$ also bends the LLs, although all in the same direction. Either at the ribbon edges or near a potential step, the eigenstates are also modified\cite{brey2006,brey2006_1,marconcini2010}. Here, an analytical expression for $\varphi_{n,k}(x)$ does not exist, thus, the solutions of the Hamiltonian in Eq. (\ref{eq:dirac4}) must be computed numerically. In these cases, the total electronic wave function $\varphi_{n,k}(x,y)$ takes the name of Edge State. Following from the initial \textit{ansatz}, an Edge State is localized in the $x$ direction but delocalized in the $y$ direction.\newline

We aim to study the temporal evolution of localized wave packets. Since Edge States are plane waves in the $y$ direction, we build a localized packet through a linear combination of multiple Edge States with fixed $n$:
\begin{equation}
\varphi_{n,\alpha}(x,y)=\int dkF(k,k^0_{\alpha},\sigma_{\alpha})\varphi_{n,k}(x)e^{iky}\, .
\end{equation}
More specifically, we choose a Gaussian weight function, $F(k,k_0^{\alpha},\sigma^{\alpha})=C\exp(-\sigma_{\alpha}^2(k-k^0_{\alpha})^2)$, and set $n=0$; $C$ is a normalization factor. This ensures a Gaussian shape of the wave packet also in real space along direction $y$. The index $\alpha$ refers to the region where $\varphi_{n,\alpha}$ is initialized: here $V(x,y)$ must be translationally invariant. The values of $k^0_{\alpha}$ and $\sigma_{\alpha}$ are the central wave vector and the real space broadening of the packet, respectively. Due to the dispersion of the LLs near the edges and/or a potential barrier, the carrier assumes a group velocity $v_g$. The slope of the bands is not constant, so that it is possible for the wave packet to lose its shape in time. Previous reports on 2DEG in GaAs/AlGaAs heterostructures\cite{beggi2015,bellentani2018,bellentani2020} indicate that a Gaussian distribution preserves its shape in a more efficient way with respect to other energy distributions, such as Lorentzian or exponential.\newline

\section{RESULTS}
\label{sec:results}
In all reported simulations we set the real-space broadening of our Gaussian wave packet to $\sigma_{\alpha}=30\, nm$. Such value corresponds to an energy broadening of $\sigma^E\simeq 7.63\, meV$. Considering three standard deviations around the central energy $E^0$, this means injecting a particle with an energy interval of $\sim 45.8\, meV$. This uncertainty would be ideal for single particle experiments, since it allows to maintain a filling factor $\nu=1$ at $B=5\, T$, and also grants an adequate shape retention of the wave packet in time. We did not observe significant differences in the results with other values of $\sigma_{\alpha}\in[10,40]\,nm$. We also fixed $\delta x=\delta y=0.5\,nm$, small enough to guarantee a good discretization, but suitable for a real-space numerical simulation. The time-dependent simulations were performed through the Split-Step Fourier method adapted to graphene nanoribbons (see Appendix \ref{sec:split-step} for a detailed derivation). The time-step $\delta t$ was set such that each exponent in the Trotter-Suzuki factorization was small enough to justify the approximation. This means $\delta t V_{max}/\hbar \ll 2\pi$.\newline

\begin{figure*}[t]
\centering
\includegraphics[width=0.9\textwidth]{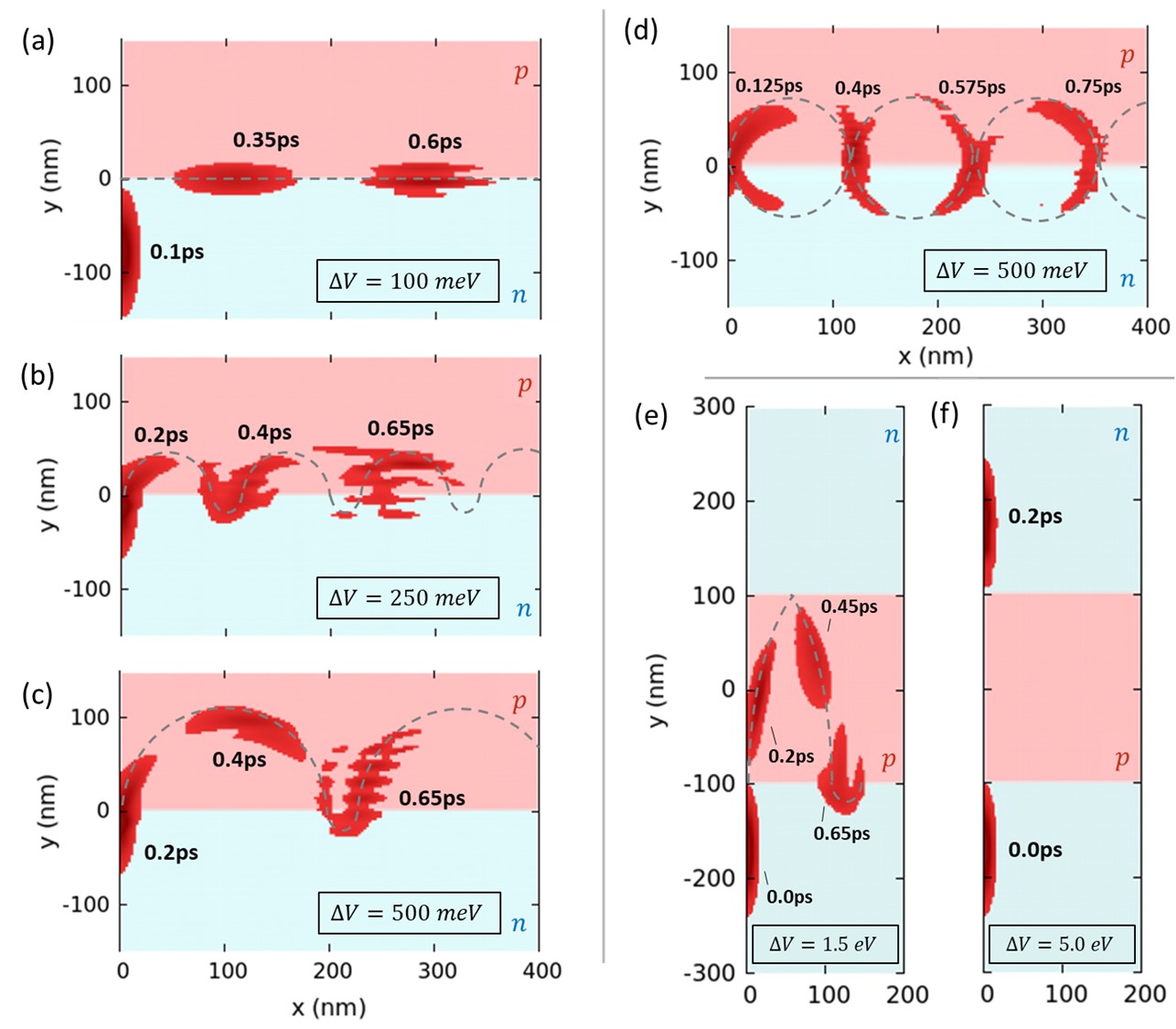}
\caption{Time-dependent simulations of a Gaussian wave packet of Edge States in armchair nanoribbons.  See also the Supplemental Material of Ref.~\cite{supplemental} for an animation of the particle dynamics. The writings in bold are the times at which each image of the packet (dark red) was taken. $\Delta V$ is the height of the potential barrier between the p region (red) and n region (blue). The dashed grey lines are a guide for the eyes. (a)-(c) Evolution along a sharp pn junction; the central energies of the packet are (a) $E^0=51.4\,meV$ and (b)-(c) $E^0=77.5\,meV$. (d) Evolution along a smooth pn junction with slope $S=0.33\, nm^{-1}$ (see Eq. (\ref{eq:fermi})) and central energy $E^0=241\,meV$. (e)-(f) Evolution along a npn junction of width $\Delta y=200\,nm$ with sharp barriers, with central energy of the wave packet $E^0=77.5\,meV$.}
\label{fig:img1}
\end{figure*}

\subsection{Evolution along a pn junction}
\label{sec:snakestates}
\subsubsection{From Edge States to Snake States}
We start the descriptions of our simulations by defining a functional regime for coherent transport of Edge States along a pn junction. Previous studies also report the physics of transport phenomena for delocalized currents in graphene pn junctions\cite{oroszlany2008,williams2011,carmier2011,barbier2012, milovanovic2014,taychatanapat2015,rickhaus2015_2,chen2016,kolasinski2017, makk2018}. In contrast, our results provide the time-resolved evolution of the carrier, taking its real-space dispersion into account, and consequently exposing the dynamics of a single quasi-particle rather than a continuous current.\newline
We call $V_p$ and $V_n$ the potential in the $p$ and $n$ regions, respectively. Figures \ref{fig:img1}(a)-(c) show the time-evolution of a Gaussian wave packet along a potential step for different values of $V_p$. Animations of these and following simulations are available as Supplemental Material at Ref.~\cite{supplemental}. The wave packet is initialized on the left edge of the ribbon away from the junction, with central energy $V_n<E^0<V_p$, and the mass term is set to $M=0$.
In Fig.~\ref{fig:img1}(a)-(c) we use a sharp potential barrier with increasing height $\Delta V=V_p-V_n$. For small $\Delta V$, the wave packet continues travelling along the junction, and is highly localized along the barrier [Fig.~\ref{fig:img1}(a)]. As $\Delta V$ increases, the carrier can penetrate more and more inside the barrier, and its path curves due to the Lorentz force. In a semi-classical picture, the curvature radius in a region with potential $V$ is the cyclotron radius\cite{barbier2012},
\begin{equation}
r_c = \frac{|\mathbf{p}|}{|q|B} = \frac{|E^0-V|}{v_F|q|B}\, ,
\end{equation}
obtained by considering the linear approximation valid near the Fermi energy, $|E^0-V|=v_F |\mathbf{p}|$, with $\mathbf{p}$ the linear momentum. If $V_n<E^0<V_p$, as is our case, the incoming carrier occupies electron-like states in the $n$ region, and hole-like states in the $p$ region. Consequently, the effective charge $q$ is negative in the $n$ region, but positive in the $p$ region. This is why the curvature of the path changes every time the packet hits the junction. Because of their specific shape, the paths in Fig.~\ref{fig:img1}(b)-(c) take the name of Snake States.

Our results confirm that both Edge States and Snake States can be present in the motion of a localized carrier in graphene. The observed transport regimes are always defined by the ratio between the cyclotron radius $r_c$ and the magnetic length $l_m$:
\begin{equation}\label{eq:ratio}
\frac{r_c}{l_m}=\frac{|E^0-V|}{v_F\sqrt{\hbar |q|B}}\, .
\end{equation}
The semi-classical picture only holds when $r_c/l_m>>1$. Notice that the ratio of Eq. (\ref{eq:ratio}) is in general different for the $p$ and $n$ zones. Table \ref{table:tab1} reports the values of $r_c/l_m$ for all the trials shown in Fig.~\ref{fig:img1}. We see that Snake States become more and more defined as the ratio increases at least in one region. Otherwise, when $r_c/l_m\ll 1$ in both regions [Fig.~\ref{fig:img1}(a)], Edge States are observed.\newline

Finally, in Fig.~\ref{fig:img1}(d) we use a smooth, Fermi-like barrier in the $y$ direction,
\begin{equation}\label{eq:fermi}
V(y)=V_n+\Delta V\frac{1}{e^{-Sy}+1}\, ,
\end{equation}
with height $\Delta V=0.5\, eV$ and slope $S= 0.33\, nm^{-1}$. The smoothness of this potential profile represents experimental conditions more realistically than the sharp case, and the smoothness parameter $S$ depends on the distance between the top gate and the device\cite{flor2022}. Although it is possible to increase the sharpness of the potential profile by positioning the top gate closer to the nanoribbon\cite{wei2017,flor2022}, it is virtually impossible to eliminate the processes that smoothen the pn junction\cite{flor2022}.\newline
Under these conditions, now the barrier presents a finite width: as a consequence, the wave packet is partially reflected every time it hits the junction. This is coherent with previous studies describing Klein tunneling in graphene pn junctions\cite{chen2016,kolasinski2017}: fixing the incident angle (here $90^{\circ}$), the reflection coefficient increases with the width of the barrier. In Fig.~\ref{fig:img1}(d) the ratio $r_c/l_m$ is similar in both regions, so that we observe similar skipping orbits on both sides of the barrier.\newline

\begin{table}[b]
\begin{ruledtabular}
\begin{tabular}{lcccccc}
Fig.~\ref{fig:img1} & (a) & (b) & (c) & (d) & (e) & (f) \\
\hline
p-zone                               & 0.386                   & 1.37                    & 3.36                    & 2.06                    & 11.3                    & 39.1                    \\
n-zone                               & 0.408                   & 0.616                   & 0.616                   & 1.92                    & 0.616                   & 0.616    \\              
\end{tabular}
\end{ruledtabular}
\caption{Values of the ratio $r_c/l_m$ for each simulation reported in Fig.~\ref{fig:img1}. The first line reports which sub-image from (a) to (f) the values refer to; p- and n-zone refer to the p region (red) and n region (blue) of the junctions.}
\label{table:tab1}
\end{table}

\subsubsection{Tunneling through a npn junction}
We also simulate npn junctions, i.e. transport in the presence of potential barriers of height $\Delta V$ and with a short length along the $y$ direction, $\Delta y$. In this section we always use sharp barriers. Figures \ref{fig:img1}(e)-(f) show two different simulations with increasing $\Delta V$ and fixed $\Delta y=0.2\,\mu m$. In both cases we set $r_c/l_m\gg 1$ in the $p$ region, as shown in Table \ref{table:tab1}. Additionally, we consider values of $r_c>\Delta y$. Figure \ref{fig:img1}(e) shows that in these conditions the wave packet cannot complete its path and is reflected back inside the $p$ zone. The carrier is reflected each time it hits the barriers: eventually it does reach the other edge of the ribbon, but in a longer time than in cases (a)-(d). Finally, Fig.~\ref{fig:img1}(f) shows that for very high $\Delta V$ the wave packet can tunnel through the barrier.\newline

We explain the behavior in Fig.~\ref{fig:img1}(f) as a process of elastic scattering between all available LLs, also called equilibration\cite{amet2014,xiang2016,zimmermann2017,wei2017}. When a particle with energy $V_n<E^0<V_p$ hits the barrier, it can scatter and partially fill the available hole-like LLs within the barrier. A higher barrier implies more available LLs: due to the slope of LLs near a potential step, the carrier will scatter further and further inside the barrier. Eventually the carrier scatters so deep inside the barrier that LLs of the second barrier become available. The slope of these LLs prevents carriers to proceed along the junction, so that the motion continues along the edge. Other than with a high $\Delta V$, this can also happen with a narrow junction, i.e. with small $\Delta y$. It is known that electrons in graphene can perfectly tunnel through potential barriers (Klein tunneling), specifically because of the presence of hole-like states within the barrier itself\cite{katsnelson2006,young2009,carmier2011,tudorovskiy2012,barbier2012}. However, we are not aware of an experimental observation of this specific phenomenon for single particles in the IQH regime in graphene.\newline

\begin{figure*}[t]
\centering
\includegraphics[width=1\textwidth]{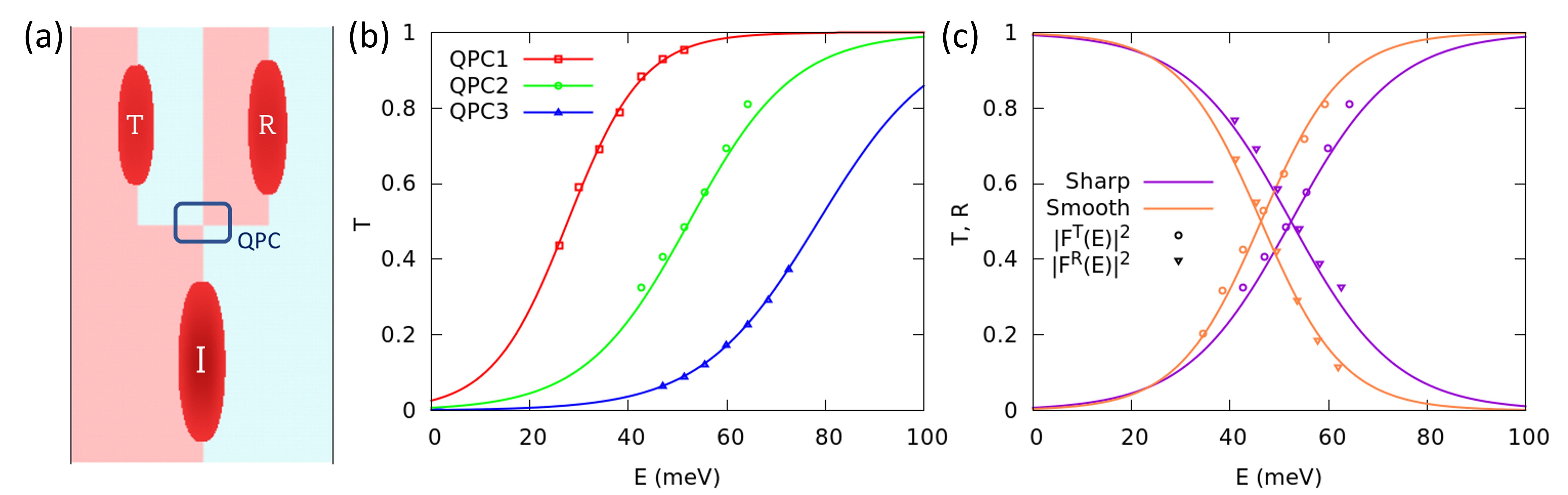}
\caption{(a) Simulation of the scattering by a QPC. The potential profile is drawn in red. `I' stands for the initial wave packet before going through the QPC; `T' and `R' and the transmitted and reflected parts, respectively. (b) Transmission curves of three different QPCs with width $L_{QPC}=5\,nm$ (QPC1), $L_{QPC}=0\,nm$ (QPC2) and $L_{QPC}=-5\,nm$ (QPC3). The points are the numerical data obtained from the projection of the transmitted wave function $\varphi^T$, the solid line is the fit. The curves were obtained for wave packets of central energy $58.2\,meV$ (QPC1), $49.8\,meV$ (QPC2) and $41.8\,meV$. (c) Transmission and reflection curves of QPC2 using a sharp barrier and a smooth barrier with slope $S=0.33\,nm^{-1}$ (see Eq. (\ref{eq:fermi})). The points are the numerical data obtained from the projection of the transmitted/reflection wave function $\varphi^{T/R}$, the solid line is the fit. The central energy of the wave packet is $49.8\,meV$.}
\label{fig:img2}
\end{figure*}

\subsection{Quantum Point Contacts}
\label{sec:QPC}
An important block used to build electronic interferometers in graphene is the Quantum Point Contact (QPC)\cite{xiang2016,zimmermann2017,ronen2021}. We simulated the behavior of QPCs in graphene using narrow constrictions of the electrostatic potential $V$. To suppress the scattering in multiple hole-like LLs we work in the Edge-State regime illustrated in the previous section, by setting the height of the potential barrier to $\Delta V = 100\,meV$; additionally, we set $M=0$. Figure \ref{fig:img2}(a) shows a QPC: when the initial packet (I) meets the constriction, the chiral nature of Edge States only allows the carrier to be transmitted along the same potential profile (T) or reflected along the other barrier (R). Previous studies on QPCs in GaAs/AlGaAs heterostructures\cite{beggi2015,beggi2015_2} proved that the reflection and transmission coefficients depend on the wave vector of each component of the wavepacket. However, in the $k$-intervals we considered for our wavepackets the band dispersion is nearly linear. As a consequence, the same empyrical formula also applies for the energies:
\begin{equation}\label{eq:QPC}
\begin{bmatrix} r(E) \\ t(E) \end{bmatrix} = \frac{1}{\sqrt{e^{\pm\alpha(E-E^{QPC})}+1}}\, .
\end{equation}
The parameters $\alpha$ and $E^{QPC}$ depend on the specific QPC: on its width $L_{QPC}$, the height of the potential $V$, and the slope of the barrier $S$. The set up of Fig.~\ref{fig:img2}(a) allows us to find the numerical values of the weights $F^{T/R}(E)$ for the transmitted/reflected components through projection on the eigenstates $\varphi_{n,k}(x)$. The transmission probability is then evaluated as 
\begin{equation}
T(E)=|t(E)|^2=\frac{|F^T(E)|^2}{|F(E)|^2}\, ,
\end{equation}
where $F(E)$ is the gaussian weight used for the initialization. The same for the reflection probability $R(E)$.\newline

Figure \ref{fig:img2}(b) shows the numerical fit for the transmission probability $T(E)$ of three different QPCs with sharp barriers. The first, QPC1, is `open': it has a width of $L_{QPC}=5\,nm$ and the potential barriers constituting the QPC do not touch each other. This is the usual conformation of a QPC which has been studied and implemented in GaAs/AlGaAs heterostructures as well\cite{beggi2015,beggi2015_2,bordone2019}. In the second one, QPC2, we set $L_{QPC}=0\,nm$: the barriers touch each other and the QPC is `closed'. In the third one, QPC3, the barriers overlap and $L_{QPC}=-5\,nm$. Surprisingly, even in the last two cases the transmission probability is non negligible. In fact, hole-like states below the closed barriers are available, and the wave packet is able to scatter through them and cross the barrier through Klein tunneling. This was not a possibility for 2DEG in gapped materials, since the carrier could only occupy states that are energetically allowed, i.e. outside of the barriers.\newline

Coherently with previous results for GaAs/AlGaAs heterostructures\cite{beggi2015,beggi2015_2}, Fig.~\ref{fig:img2}(b) shows that transmission increases with the energy. Furthermore, transmission curves shift to higher energies as the QPC closes, maintaining similar slopes. This means that although Klein tunneling helps increasing transmission, the wave packet still needs more and more energy to be transmitted through progressively closing gates, at least in the functional Edge-State regime we considered. Finally, Fig.~\ref{fig:img2}(c) compares transmission and reflection curves for a sharp barrier and a smooth barrier with slope $S= 0.33\, nm^{-1}$. Both curves are obtained for QPC2. We see that a smoother barrier not only changes the slope of the curves but also shifts the value of $E_0$ to lower energies. As a consequence, components with lower energies are able to transmit through the QPC: the wave packet behaves as if the effective width of the QPC were larger.\newline

\begin{figure}[b]
\centering
\includegraphics[width=0.35\textwidth]{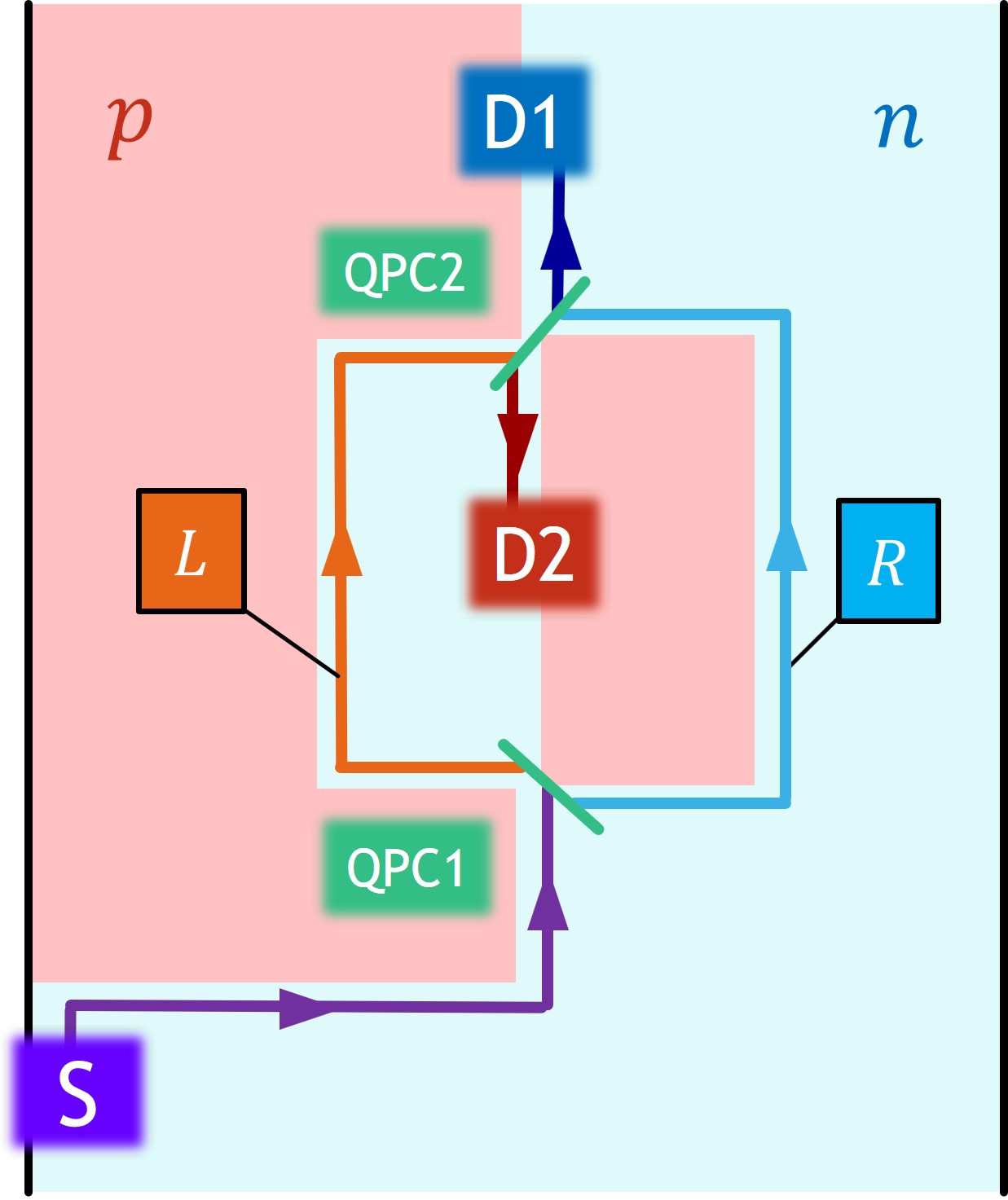}
\caption{Schematic representation of the first MZI simulated, made with QPCs. The potential profile is drawn in red. S stands for Source; QPC1/2 are the Quantum Point Contacts at the ends of the interferometer; D1/2 are the Detectors; L and R stand for the left and right arms of the interferometers, respectively. See also the Supplemental Material of Ref.~\cite{supplemental} for video animations of the wave packet propagating for three different MZI areas.}
\label{fig:img3}
\end{figure}

\begin{figure*}[t]
\centering
\includegraphics[width=1\textwidth]{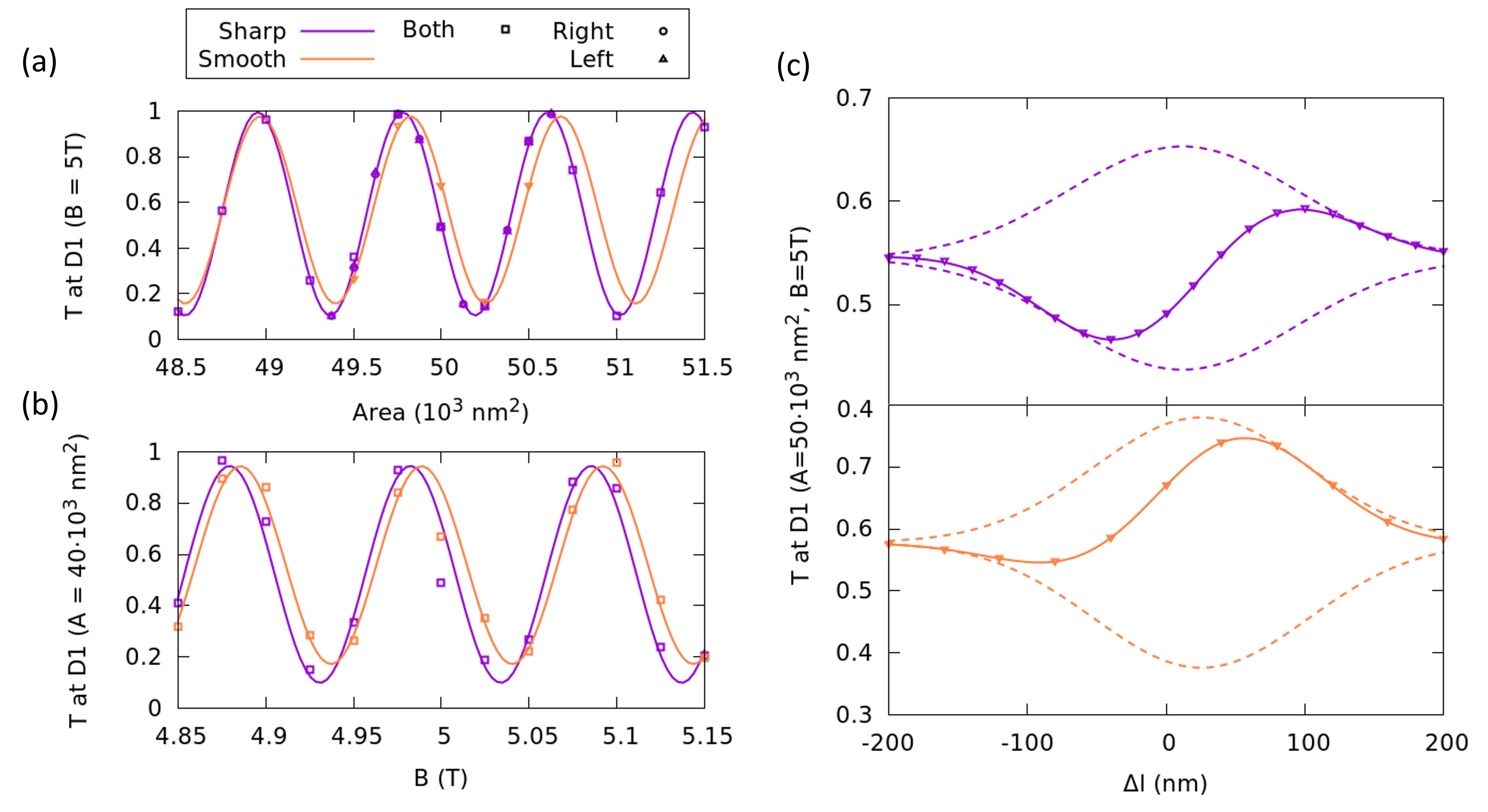}
\caption{Curves of the transmission probability at detector D1 of the MZI shown in Fig.~\ref{fig:img3}. The points are the numerical data, the solid line is the fit. The curves obtained with a sharp barrier and a smooth barrier of slope $S=0.33\,nm^{-1}$ (see Eq. (\ref{eq:fermi})) are reported in purple and orange respectively. (a) Aharonov-Bohm oscillations at fixed magnetic field $B=5\,T$ and varying the area between the arms of the interferometers. The area was modified either by changing the length of both arms (`Both') or one at a time (`Left' and `Right') (b) Aharonov-Bohm oscillations at fixed area $A=40\cdot10^3\,nm^2$ and varying the magnetic field. (c) Phase averaging curves obtained by fixing both the magnetic field $B=5\,T$ and the area $A=50\cdot 10^3\,nm^2$, and varying the path difference $\Delta l$. `G. Mod.' stands for Gaussian Modulation.}\label{fig:img4}
\end{figure*}

\subsection{MZI with Quantum Point Contacts}
\label{sec:MZI1}
We simulate a first type of MZI using two QPCs. Figure \ref{fig:img3} shows a schematic representation of our interferometer. The electronic wave packet is emitted at the Source (S). The first QPC separates the wave packet into left and right paths (L/R), which constitute the arms of the interferometer. The two parts of the wave packet $\varphi^R$ and $\varphi^L$ later reunite at the second QPC, and can interfere. Eventually, the carrier can be detected either at detector D1 or D2. Our group has studied similar single-particle interferometers in GaAs/AlGaAs heterostructures as well\cite{beggi2015,beggi2015_2,bordone2019}. For graphene however, several studies\cite{abanin2007,xiang2016,marguerite2019,moreau2021,kumar2022} show how the presence of ribbon edges may be detrimental for the implementation of such interferometers, due to back-scattering under the gates. To reduce this phenomenon, we keep the arms of the interferometer away from the edges. We believe that the potential profile we considered can at least partially bypass these backscattering process, in light of various experiments in edgeless geometries that allow for extremely clean measurements\cite{zeng2019,polshyn2018}. We work again in the Edge-State regime, by setting a barrier of $\Delta V=100\,meV$, and setting $M=0$. We select the `closed' QPC2 from the previous section ($L_{QPC}=0\,nm$). Then, we initialize the wave packet in order to have an approximate $43-57\%$ splitting (sharp barrier) and $44-56\%$ splitting (smooth barrier) after the first QPC. This way, being close to the ideal case of $50-50\%$ splitting, we ensure a non negligible interference at the end of the interferometer.\newline

In Fig.~\ref{fig:img4} we show the transmission probability $T$ through the second QPC obtained from our simulations. Notice that $T$ is equivalent to the probability of detecting the carrier at D1. For each trial we consider both cases with a sharp barrier and a smooth one with slope $S= 0.33\, nm^{-1}$. Figure \ref{fig:img4}(a) and (b) show Aharanov-Bohm oscillations when varying the area of the MZI and the magnetic field, respectively. The deviation from the expected period is around $3.5\%$ when varying the area in presence of smooth barriers; in all other cases it remains below $0.5\%$. A reason behind the discrepancy may lie in the spatial spread of the carrier, as well as in the central position $x_{R/L}$ of the wavepackets along the arms of the interferometers, which in general may add an additional phase. However, note that in all cases the considered flux is approximately 50 to 60 times the unit flux $\Phi_0=e/\hbar$, so that these errors are still negligible. The overall visibility is very high, reaching an approximate value of $81\%$ for the sharp barrier, and $73\%$ for the smooth barrier. Taking into account the non-ideal splitting of the wave packet into the two paths of the interferometer, these values correspond to respectively $98.8\%$ and $95\%$ the maximal obtainable visibilities. Under this aspect, our results are coherent with previous experimental measurements of interferometers in graphene, which show a much higher visibility with respect to 2DEG in gapped semiconductors\cite{wei2017,jo2021}.\newline

Aharonov-Bohm oscillations have been observed in multiple experiments with interferometers in graphene working with delocalized currents\cite{xiang2016,wei2017,ronen2021,morikawa2015,johnson2021,jo2021,deprez2021}. However, the broad energy distribution of our localized wave packet allows us to show an additional phenomenon, detrimental to the visibility, called phase averaging\cite{chung2005,giovannetti2008}. This effect is shown in Fig.~\ref{fig:img4}(c): here, we maintain a constant flux $\Phi$ through the arms of the interferometers by fixing both the area and magnetic field. We change the length of the left and right paths of the interferometer separately, thus introducing an optical path difference $\Delta l$. For delocalized currents, this would just mean a relative phase of $\Delta\theta=(2\pi\Delta l/L+\theta_0)$. For the case of localized particles instead, we need to take account of how much $\varphi^R$ and $\varphi^L$ overlap when they reunite. In fact, when $|\Delta l|$ is close to zero, $\varphi^R$ and $\varphi^L$ arrive at the second QPC simultaneously and overlap completely. Instead, a value of $|\Delta l|\gg 0$ introduces a significant delay between the two, thus reducing the overlap between $\varphi^R$ and $\varphi^L$, and consequently the interferometer's visibility. Eventually, when $\Delta l$ is too large, they arrive at the second QPC separately and no longer interfere. The overall modulation $F^{avg}$ of the transmission coefficient $T$ comes from the convolution of the weight functions of $\varphi^R$ and $\varphi^L$. If we consider both of them to maintain a Gaussian shape after the splitting, this reads:
\begin{equation}
F^{avg}(\Delta l)=(F^R*F^L)(\Delta l) = Ce^{-\frac{(\Delta l - \Delta l_0)^2}{4(\sigma_R^2+\sigma_L^2)}}
\end{equation}
where $\sigma_{R/L}$ are the real space broadening in the $y$ direction; $C$ is a normalization factor. Notice the presence of an additional path difference $\Delta l_0$, which is generally different from zero. This additional relative phase may be due to the centers $x_{R/L}$ of each packet being generally displaced from the center of the barrier, an effect which modifies their effective paths. In Fig.~\ref{fig:img4}(c) are drawn the fits of our data with the equation
\begin{equation}
T(\Delta l)=T_0+F^{avg}(\Delta l)\cdot \Delta T \sin\left(\frac{2\pi\Delta l}{L}+\theta_0\right)\, ,
\end{equation}
which agree well with our results. Note that $\Delta l_0$ and $\theta_0$ are both relative phases. However, $\Delta l_0$ appears due to the localization of the packets, and influences the superposition of $\varphi^{R/L}$. Instead, $\theta_0$ is a relative phase in the interference, and would also be present in the case of delocalized currents.\newline
The important thing to notice from Fig.~\ref{fig:img4}(c) is the shift of the Gaussian modulation between the cases of sharp and smooth barriers: these are respectively centered around $\Delta l_0^{sharp}\simeq12\,nm$ and $\Delta l_0^{smooth}\simeq25\,nm$. In fact, a smooth potential implies a lower slope of the bands and a smaller group velocity: $v_g=-\frac{1}{\hbar}\frac{\partial E(k)}{\partial k}$ ranges from $v_g^{sharp}=7.25\cdot10^5\, m/s$ to $v_g^{smooth}=6.97\cdot10^5\, m/s$. Thus, the intrinsic delay increases in the case of smooth barriers.\newline

\begin{figure}[b]
\centering
\includegraphics[width=0.35\textwidth]{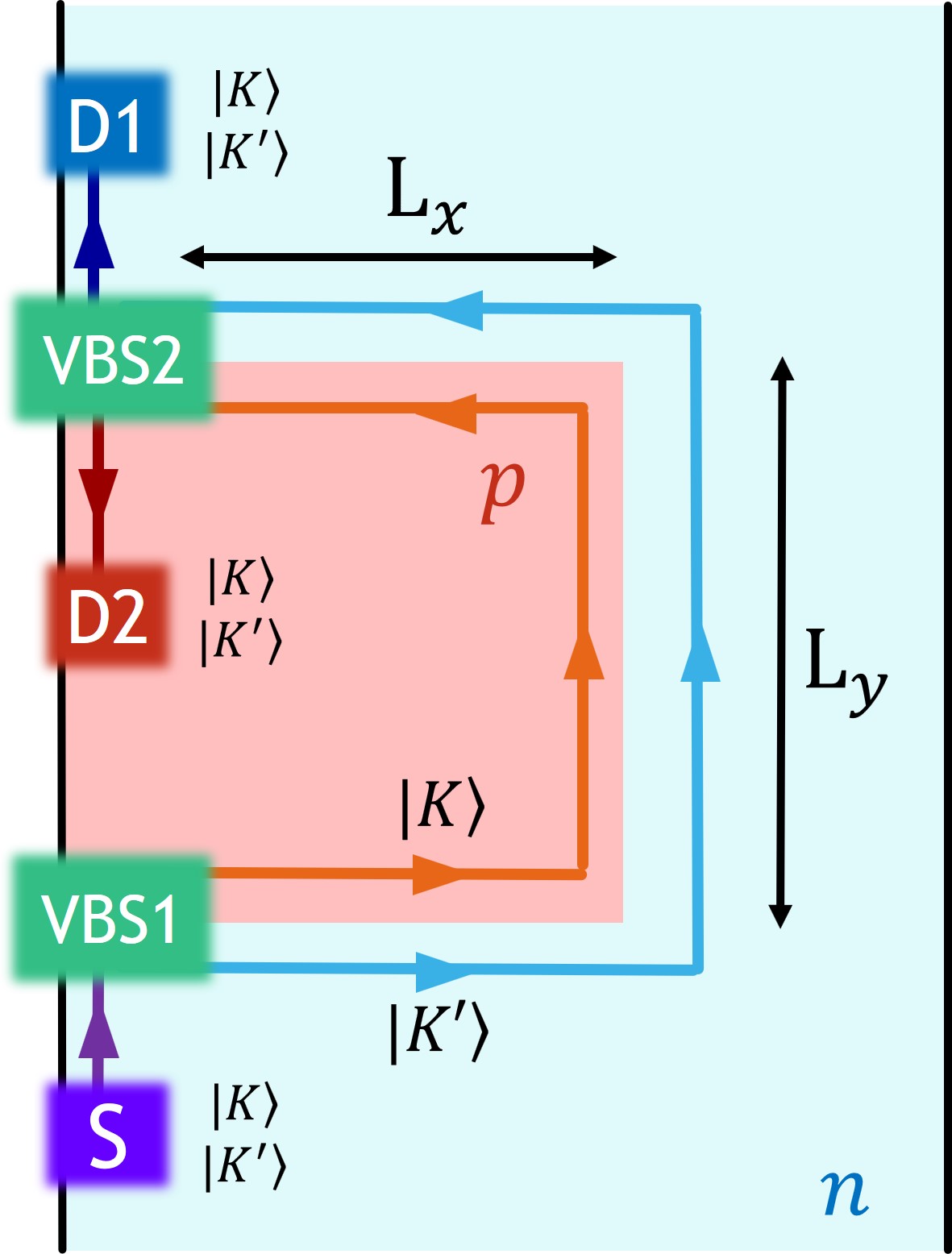}
\caption{Schematic representation of the second MZI, made with VBSs. The potential profile of the gate is drawn in red. S stands for Source; VBS1/2 are the Valley Beam Splitters at the ends of the interferometer; D1/2 are the Detectors; the $K$ and $K'$ arms of the interferometer are represented as orange and blue lines, respectively. See also the Supplemental Material of Ref.~\cite{supplemental} for video animations of the wave packet propagating for three different intensities of the magnetic field.}
\label{fig:img5}
\end{figure}

\begin{figure*}[t]
\centering
\includegraphics[width=1\textwidth]{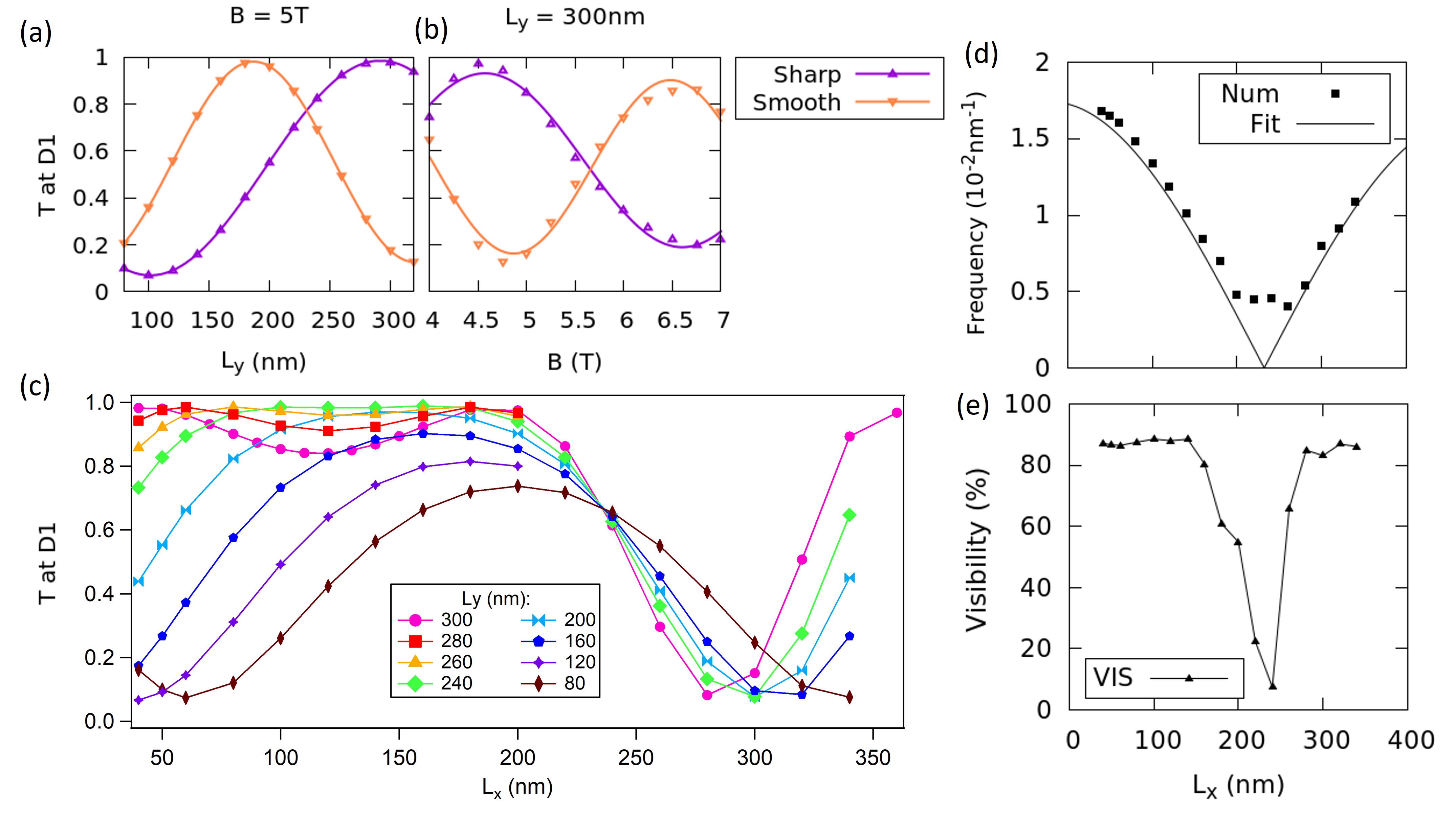}
\caption{(a)-(c) Curves of the transmission probability at detector D1 of the MZI shown in Fig.~\ref{fig:img5}. The points are the numerical data, the solid line is the fit. `Sharp' stands for a sharp barrier; `Smooth' for a barrier of slope $S=0.33\,nm^{-1}$. (a) Aharonov-Bohm oscillations at fixed magnetic field $B=5\,T$ and dimension along $x$ of the gate, $L_x=50\,nm$, while varying $L_y$. (b) Aharonov-Bohm oscillations at fixed gate dimensions $L_x$ and $L_y$ with varying magnetic field $B$. (c) Transmission curves with a sharp barrier for different values of $L_y$, observed by varying $L_x$ at fixed $B=5\,T$. (d)-(e) Respectively frequence and visibility of the Aharonov-Bohm oscillations obtained with a sharp barrier for different values of $L_x$, observed by varying $L_y$ at fixed $B=5\,T$.}
\label{fig:img6}
\end{figure*}

\begin{table*}[t]
\begin{ruledtabular}
\begin{tabular}{l|r|rrrrr|rrrr}
$E_g (meV$)                 & 10    & 20    & 20    & 20     & 20    & 20    & 40    & 40    & 40    & 40     \\ \hline
S ($nm^{-1}$) & Sharp & Sharp & 0.50     & 0.33 & 0.25   & 0.10     & Sharp & Sharp & Sharp & 0.33 \\ \hline
$E_0 (meV$)                 & 48.84 & 48.84 & 48.84 & 48.84  & 48.84 & 48.84 & 42.71 & 48.84 & 56.78 & 48.84  \\ \hline
$T^{VBS} (\%)$                   & 46.83 & 43.66 & 42.69 & 41.78  & 40.78 & 34.82 & 37.19 & 37.41 & 37.32 & 33.95  \\
$R^{VBS} (\%)$                   & 52.97 & 55.92 & 56.79 & 57.55  & 58.32 & 63.05 & 61.56 & 61.52 & 61.58 & 64.73 
\end{tabular}
\end{ruledtabular}
\caption{Values of the transmission ($T^{VBS}$) and reflection ($R^{VBS}$) probabilities of different VBSs. $E_g$ is the energy gap between $LL_0^{+/-}$; S is the slope of the barrier (see Eq. (\ref{eq:fermi})), `Sharp' being a sharp barrier; $E_0$ is the central energy of the simulated wavepacket.}
\label{table:tab2}
\end{table*}

\subsection{MZI with Valley Beam Splitters}
\label{sec:MZI2}
Finally, we simulate the evolution of Gaussian wavepackets in a MZI built with Valley Beam Splitters (VBS). For these simulations we still consider the Edge-State regime; additionally, we induce a mass gap by setting $M\neq 0$. This way, the valley degeneracy of $LL_0^{+/-}$ is lifted. At the edge of the ribbon, $LL_0^{+/-}$ have opposite curvatures and carry ESs with opposite chirality. Along a potential barrier instead, $LL_0^{+/-}$ bend the same way: as a consequence, electron-like and hole-like ESs propagate in the same direction. Under these conditions, VBSs appear where the potential barrier meets the edge of the ribbon\cite{wei2017,jo2021}. This phenomenon occurs at every intersection between the potential gates and the physical edge of the device, as described and measured in Ref.~\cite{wei2017}. There, VBSs were created at the border separating two regions of different filling factors, which in turn were created by suitable top gates. In fact, when the Fermi level crosses multiple channels it allows for equilibration processes between edge channels, but only if they are running along the physical edges. Our setup is shown in Fig.~\ref{fig:img5}: the electron is initialized at source S at the edge of the ribbon. Due to the nature of the armchair edge, the electronic state is in a superposition of valley $K$ and $K'$\cite{brey2006,marconcini2010}. Once the carrier evolves and meets the potential barrier (at VBS1), it can scatter elastically between the available $LL_0^{+/-}$ states, and consequently splits in the two components $\Psi^K$ and $\Psi^{K'}$, belonging to the $K$ and $K'$ valleys, respectively. Consequently, $\Psi^K$ and $\Psi^{K'}$ travel along two distinct paths, until they reunite at VBS2, i.e. where the potential profile meets the edge again. Here $\Psi^K$ and $\Psi^{K'}$ interfere, and the carrier can be detected either at D1 or D2.\newline
The set up shown in Fig.~\ref{fig:img5} is similar to several experimental implementations of MZIs with VBSs \cite{morikawa2015,wei2017,jo2021,assouline2021}. However, these devices worked along the full nanoribbon width. Instead, we decided to set both VBSs on the same edge by setting a potential profile which defines a region that does not reach the opposite edge of the ribbon. Our simulations show that two fully functional VBSs are formed in this case as well. With this set up, we are able to change the length of the interferometer's arms by manipulating the dimensions $L_x$ and $L_y$ of the gate, as defined in Fig.~\ref{fig:img5}.\newline

We first characterize the behavior of a VBS in presence of a localized wavepacket. Table \ref{table:tab2} shows the splitting percentages of the packet into the transmitted $K$ component ($T^{VBS}$) and reflected $K'$ component ($R^{VBS}$). Notice that the percentages do not sum up to $100\%$ due to a small splitting into the other LLs. In contrast with the case of QPCs, transmission and reflection of the wave packet do not depend on its initialization energy $E^0$. Instead, $T^{VBS}$ and $R^{VBS}$ only depend on the value of the mass term $M$ and the slope of the barrier $S$. To be more specific, $T^{VBS}$ decreases as $E_g=2M$ increases and as the potential gets smoother: in both cases, this happens because the separation between $LL_0^{+/-}$ increases. The limit of $50-50\%$ would be restored when $M=0$ and degeneracy is not lifted. Overall, our results are coherent with what has been reported in Ref.~\cite{flor2022}, which showed that the behavior of MZIs with VBSs only depends on the electrostatic setup of the device.\newline

Then, we measure the probability of detecting the carrier at D1, that is, the overall transmission probability $T$ through the interferometer. For these trials we select a case with $T^{VBS}\simeq 44\%$, $R^{VBS}\simeq 56\%$ for the sharp barrier and $T^{VBS}\simeq 42\%$, $R^{VBS}\simeq 58\%$ for the smooth barrier. Figure \ref{fig:img6}(a) shows Aharonov-Bohm oscillations as a function of the gate dimension $L_y$ and fixed $L_x$ and magnetic field $B$. The period $L_y^T$ of the oscillations tells us the mean value of the distance between the channels throughout the whole interferometer: $\delta^{sharp}=(2.169\pm0.003)\,nm$ and $\delta^{smooth}=(3.084\pm0.008)\,nm$. Coherently with the results in Table \ref{table:tab2}, $\delta^{sharp}<\delta^{smooth}$ because the separation of $LL_0^{+/-}$ increases for smooth barriers. Note that these $\delta$'s are much smaller than the values reported in literature for delocalized currents\cite{wei2017,jo2021,flor2022}, which are of the order of $\delta\in[50,110]\, nm$. We relate this discrepancy to the specific regime we simulate, suitable instead for localized particles. However the values of $\delta$ we found are coherent with the results of a recent study in a similar system\cite{mirzakhani2022}. We were also able to obtain a separation as large as $\delta\sim 12\, nm$ as well, however this implied a value of $T^{VBS}<5\%$.\newline
Figure \ref{fig:img6}(b) shows Aharonov-Bohm oscillations as a function of the magnetic field $B$. Changing $B$ modifies the LL dispersion, although slightly, thus changing the separation of $LL_0^{+/-}$ along the barrier. This may modify the behavior of the first VBS, which would explain the slight discrepancy between the numerical data and the fit.\newline

Figure \ref{fig:img6}(c) shows the overall transmission probability $T$ as a function of the horizontal gate dimension, $L_x$, for various values of $L_y$. The non-trivial behavior of $T$ along the $L_x$ axis is a consequence of how the separation $\delta$ between the channels varies along the interferometer's arms. To be more specific, the distance $\delta_y$ between the spatial centers of $\Psi^K$ and $\Psi^{K'}$ along $L_y$ oscillates as a function of the length of $L_x$. We find $\delta_y$ to have a periodicity of $L_x^T\simeq 992\,nm$, and observe that the Edge Channels cross for certain values of $L_x$. Figure \ref{fig:img6}(d) shows the frequencies of the Aharonov-Bohm oscillations along $L_y$ for different values of $L_x$. The graph compares the numerical data with the expected values, $f=eB|\delta_y|/\hbar$, obtained from a previous fit of $\delta_y$. The data differs from the fit only when the frequencies are small: in this interval, $\delta_y$ is close to zero, and we believe the energy broadening of the packets may affect the oscillations. A further confirmation of this comes from Fig.~\ref{fig:img6}(e), in which we observe a significant drop in the visibility in the same interval.\newline

In other words, we suggest that the Edge Channels in graphene MZIs do not have a trivial shape. On the contrary, they oscillate when propagating along the $x$ direction, that is, along a zigzag direction orthogonal to the armchair edges of the ribbon. Instead, the same Edge Channels do not oscillate in the armchair direction. The oscillation along $x$ is of the order of few nanometers, which is why its effects were negligible in previous implementations\cite{wei2017,jo2021} with a mean $\delta\in[50,110]\, nm$. On the contrary, our interferometer works with a much smaller separation between the channels, and we are able to observe crossing between Edge Channels.\newline

\section{CONCLUSIONS}\label{sec:conclusions}
The time-dependent simulations we reported show the functional regimes for the transport of localized wave packets of Edge States in MZIs in graphene. Up to now there have been several studies on the behavior of such interferometers with delocalized currents\cite{wei2017,jo2021,assouline2021,xiang2016}. Instead, by taking into account the real-space dispersion of the carrier, we were able to study the specific behaviour of localized quasi-particles in the scope of possible future single- and few-particle implementations of the same devices\cite{kotilahti2021, edlbauer2022}. We solved the exact time-dependent Schr\"odinger equation in order to observe the real time evolution of the particle in real space and describe its dynamics. We first showed that it is possible to control the transport regime of single particles along graphene pn junctions through the interplay between the energy of the injected carrier and the height of the junction itself. More specifically, we showed that the tuning of the cyclotron radius of the carrier in both regions of the junction makes it possible to either observe semiclassical Snake State trajectories or Edge State behavior. We also provided proof of the same phenomena that have been previously observed for delocalized currents, such as Klein tunneling through a finite potential barrier and the related dependence of the reflection coefficient on the smoothness of the barrier\cite{chen2016,kolasinski2017}.\newline
We then studied the behavior of single particle transport in two kinds of MZIs. The first MZI is built with QPCs: these interferometers have been widely studied in systems of GaAs/AlGaAs heterostructures\cite{bordone2019}. In graphene, the same devices are subject to detrimental effects due to channel scattering at the physical edges of the device\cite{xiang2016,kumar2022}. However, we showed that it is possible to suppress such effects with the use of an appropriate gate shape. Thus, we observed clear, high-visibility Aharonov-Bohm oscillations. Additionally, we have showed the occurrence of phase averaging\cite{giovannetti2008,chung2005} caused by the characteristic energy spread of localized carriers. This phenomenon is not present for a delocalized current, but it will become crucial in devising future single- and few-particle interferometers, since it is detrimental to the visibility.\newline
The second kind of MZI is built with VBSs and reproduces those used in recent experimental studies\cite{wei2017,jo2021,assouline2021}. In this paper, we propose a new geometry for this device, which allows to design the length of the arms of the interferometer regardless of the width on the graphene nanoribbon. This way, it allows for additional freedom in engineering the devices. That said, while the scattering processes of QPCs strongly depend on the energy of the injected carrier, the transmission and reflection coefficients of VBSs only depend on the electrostatic setup of the device\cite{flor2022}. To obtain a splitting close to the ideal $50-50\%$ in presence of localized carriers, we have found that one needs to work with clean graphene samples with small energy gaps and potential barriers as sharp as possible. Overall, we have studied a regime that is different to those previously observed: while in Ref.~\cite{jo2021} the channel separation is as large as $\delta \sim 110\,nm$, we have considered a situation with $\delta$ of the order of a few nanometers. By doing so, we could see that this distance is not constant along the zigzag direction, and influences the interference non-trivially. To our knowledge, this phenomenon has not been previously reported. Our results imply the presence of non trivial transport processes that occur along a pn junction, with non negligible effects in the specific regime considered for single particle transport.

\appendix

\section{The Split-Step Method in graphene}
\label{sec:split-step}
Once the initial state of the system $\varphi(x,y;t_0)$ is defined, its temporal evolution is obtained by integrating the time-dependend Schr\"odinger equation. To do so, we define an evolution operator:
\begin{equation}
\hat{U}(\Delta t)=e^{-\frac{i}{\hbar}\hat{H}\Delta t}\, ,
\end{equation}
where $\hat{H}$ is the Hamiltonian in Eq. (\ref{eq:dirac3}), independent of time; $\Delta t$ is the total evolution time-step. To apply the evolution numerically, we exploit the Split-Step Fourier method through the Trotter-Suzuki factorization\cite{kramer2010,chaves2015,grasselli2015,grasselli2016} for infinitesimal $\delta t=\Delta t / N\rightarrow 0$. We rewrite $\hat{H}$ from Eq. (\ref{eq:dirac3}) as a sum of multiple non-commuting terms:
\begin{align}\label{eq:splitsum}
\hat{H} = \hat{V}(\left|x\right|,y)+\hat{T}_x(k_x)+\hat{T}_y(k_y)+\hat{V}_B(\left|x\right|)\, ,
\end{align}
\begin{equation}\label{eq:evol}
\begin{cases}
\hat{V}(|x|,y)=V(|x|,y)\cdot \mathbb{1}^{AB}+M\cdot \sigma_z^{AB} \\
\hat{T}_{x/y}=-\hbar v_F k_{x/y}\cdot \sigma_{x/y}^{AB} \\
\hat{V}_B(|x|)=-\frac{\hbar v_F}{l_m^2}|x|\cdot \sigma_y^{AB}
\end{cases}\, .
\end{equation}
In Eq. (\ref{eq:evol}), $\sigma^{AB}_{x/y/z}$ are the Pauli matrices on the sublattice basis; $\mathbb{1}^{AB}$ is the identity matrix. The operator $\hat{V}$ is a block-diagonal potential term which includes the mass term $M$. $\hat{T}_{x/y}$ are off-diagonal kinetic operators linear in $k_x$ and $k_y$, respectively. Finally, $\hat{V}_B$ is the magnetic potential, non-diagonal as well: it is linear both in $|x|$ and $B$. The Trotter-Suzuki factorization provides the following approximation for $\hat{U}(\Delta t)$:
\begin{align}\label{eq:trotter}
[e^{-\frac{i}{\hbar}\hat{H}\delta t}]^N = &\, e^{-\frac{i}{\hbar} \hat{V}\frac{\delta t}{2}}[  e^{-\frac{i}{\hbar}\hat{T}_x\delta t}e^{-\frac{i}{\hbar}\hat{T}_y\delta t} \nonumber \\
& \qquad e^{-\frac{i}{\hbar}\hat{V}_B\delta t}e^{-\frac{i}{\hbar}\hat{V}\delta t}]^Ne^{+\frac{i}{\hbar}\hat{V}\frac{\delta t}{2}}\, .
\end{align}
The addition of the two-dimensional Fourier transform $F$ allows us to switch from real space $(x,y)$ to the reciprocal space $(k_x,k_y)$, where the locality of $\hat{T}_{x/y}$ can be exploited. Vice versa for the antitransform $F^{-1}$. The final evolution operator reads:
\begin{align}\label{eq:split}
\hat{U}(\Delta t) = &\, e^{-\frac{i}{\hbar} \hat{V}\frac{\delta t}{2}}[F^{-1}  e^{-\frac{i}{\hbar}\hat{T}_x\delta t}e^{-\frac{i}{\hbar}\hat{T}_y\delta t} \nonumber \\
& \qquad F e^{-\frac{i}{\hbar}\hat{V}_B\delta t}e^{-\frac{i}{\hbar}\hat{V}\delta t}]^Ne^{+\frac{i}{\hbar}\hat{V}\frac{\delta t}{2}}\, .
\end{align}
Since $\hat{V}$ is block-diagonal, its exponentiation is trivial. On the contrary, the exponential of the non-diagonal $\hat{T}_{x/y}$ and $\hat{V}_B$ leads back to the exponentiation of $\sigma_{x/y}$\cite{chaves2015}: the results are $2\times 2$ block-operators with all elements different from zero. The main consequence of the shape of the operators in Eq. (\ref{eq:split}) is that the evolutions of the two components of the spinor $\varphi$ are coupled: so, they must be carried out at the same time.\newline

The definition of $\varphi$ in the previous Section requires a mirrored space in the $x$ direction. Thus, a real ribbon defined in $[0,L_x]\times[-L_y/2,L_y/2]$ corresponds to an actual simulated space defined in $[-L_x,L_x]\times[-L_y/2,L_y/2]$. The Fourier Transforms needed for the Split-Step method are performed through the FFT algorithm, allowing us to impose periodic conditions in both the $x$ and $y$ directions. However, the simulated space is still a nanoribbon of finite width, because of how we defined the problem for $\varphi$.\newline


\bibliography{bibliography.bib}

\begin{thebibliography}{75}%
\makeatletter
\providecommand \@ifxundefined [1]{%
 \@ifx{#1\undefined}
}%
\providecommand \@ifnum [1]{%
 \ifnum #1\expandafter \@firstoftwo
 \else \expandafter \@secondoftwo
 \fi
}%
\providecommand \@ifx [1]{%
 \ifx #1\expandafter \@firstoftwo
 \else \expandafter \@secondoftwo
 \fi
}%
\providecommand \natexlab [1]{#1}%
\providecommand \enquote  [1]{``#1''}%
\providecommand \bibnamefont  [1]{#1}%
\providecommand \bibfnamefont [1]{#1}%
\providecommand \citenamefont [1]{#1}%
\providecommand \href@noop [0]{\@secondoftwo}%
\providecommand \href [0]{\begingroup \@sanitize@url \@href}%
\providecommand \@href[1]{\@@startlink{#1}\@@href}%
\providecommand \@@href[1]{\endgroup#1\@@endlink}%
\providecommand \@sanitize@url [0]{\catcode `\\12\catcode `\$12\catcode
  `\&12\catcode `\#12\catcode `\^12\catcode `\_12\catcode `\%12\relax}%
\providecommand \@@startlink[1]{}%
\providecommand \@@endlink[0]{}%
\providecommand \url  [0]{\begingroup\@sanitize@url \@url }%
\providecommand \@url [1]{\endgroup\@href {#1}{\urlprefix }}%
\providecommand \urlprefix  [0]{URL }%
\providecommand \Eprint [0]{\href }%
\providecommand \doibase [0]{https://doi.org/}%
\providecommand \selectlanguage [0]{\@gobble}%
\providecommand \bibinfo  [0]{\@secondoftwo}%
\providecommand \bibfield  [0]{\@secondoftwo}%
\providecommand \translation [1]{[#1]}%
\providecommand \BibitemOpen [0]{}%
\providecommand \bibitemStop [0]{}%
\providecommand \bibitemNoStop [0]{.\EOS\space}%
\providecommand \EOS [0]{\spacefactor3000\relax}%
\providecommand \BibitemShut  [1]{\csname bibitem#1\endcsname}%
\let\auto@bib@innerbib\@empty
\bibitem [{\citenamefont {Novoselov}\ \emph {et~al.}(2005)\citenamefont
  {Novoselov}, \citenamefont {Geim}, \citenamefont {Morozov}, \citenamefont
  {Jiang}, \citenamefont {Katsnelson}, \citenamefont {Grigorieva},
  \citenamefont {Dubonos},\ and\ \citenamefont {Firsov}}]{novoselov2005}%
  \BibitemOpen
  \bibfield  {author} {\bibinfo {author} {\bibfnamefont {K.~S.}\ \bibnamefont
  {Novoselov}}, \bibinfo {author} {\bibfnamefont {A.~K.}\ \bibnamefont {Geim}},
  \bibinfo {author} {\bibfnamefont {S.~V.}\ \bibnamefont {Morozov}}, \bibinfo
  {author} {\bibfnamefont {D.}~\bibnamefont {Jiang}}, \bibinfo {author}
  {\bibfnamefont {M.~I.}\ \bibnamefont {Katsnelson}}, \bibinfo {author}
  {\bibfnamefont {I.~V.}\ \bibnamefont {Grigorieva}}, \bibinfo {author}
  {\bibfnamefont {S.~V.}\ \bibnamefont {Dubonos}},\ and\ \bibinfo {author}
  {\bibfnamefont {A.~A.}\ \bibnamefont {Firsov}},\ }\href@noop {} {\bibfield
  {journal} {\bibinfo  {journal} {Nature}\ }\textbf {\bibinfo {volume} {438}},\
  \bibinfo {pages} {197} (\bibinfo {year} {2005})}\BibitemShut {NoStop}%
\bibitem [{\citenamefont {Zhang}\ \emph {et~al.}(2005)\citenamefont {Zhang},
  \citenamefont {Tan}, \citenamefont {Stormer},\ and\ \citenamefont
  {Kim}}]{zhang2005}%
  \BibitemOpen
  \bibfield  {author} {\bibinfo {author} {\bibfnamefont {Y.}~\bibnamefont
  {Zhang}}, \bibinfo {author} {\bibfnamefont {Y.-W.}\ \bibnamefont {Tan}},
  \bibinfo {author} {\bibfnamefont {H.}~\bibnamefont {Stormer}},\ and\ \bibinfo
  {author} {\bibfnamefont {P.}~\bibnamefont {Kim}},\ }\href
  {https://doi.org/10.1038/nature04235} {\bibfield  {journal} {\bibinfo
  {journal} {Nature}\ }\textbf {\bibinfo {volume} {438}},\ \bibinfo {pages}
  {201} (\bibinfo {year} {2005})}\BibitemShut {NoStop}%
\bibitem [{\citenamefont {Ho}\ \emph {et~al.}(2008)\citenamefont {Ho},
  \citenamefont {Lai}, \citenamefont {Chiu},\ and\ \citenamefont
  {Lin}}]{ho2008}%
  \BibitemOpen
  \bibfield  {author} {\bibinfo {author} {\bibfnamefont {J.}~\bibnamefont
  {Ho}}, \bibinfo {author} {\bibfnamefont {Y.}~\bibnamefont {Lai}}, \bibinfo
  {author} {\bibfnamefont {Y.-H.}\ \bibnamefont {Chiu}},\ and\ \bibinfo
  {author} {\bibfnamefont {M.}~\bibnamefont {Lin}},\ }\href
  {https://doi.org/10.1016/j.physe.2007.10.065} {\bibfield  {journal} {\bibinfo
   {journal} {Physica E-low-dimensional Systems $\&$ Nanostructures - PHYSICA
  E}\ }\textbf {\bibinfo {volume} {40}},\ \bibinfo {pages} {1722} (\bibinfo
  {year} {2008})}\BibitemShut {NoStop}%
\bibitem [{\citenamefont {Wu}\ \emph {et~al.}(2009)\citenamefont {Wu},
  \citenamefont {Hu}, \citenamefont {Ruan}, \citenamefont {Madiomanana},
  \citenamefont {Hankinson}, \citenamefont {Sprinkle}, \citenamefont {Berger},\
  and\ \citenamefont {de~Heer}}]{wu2009}%
  \BibitemOpen
  \bibfield  {author} {\bibinfo {author} {\bibfnamefont {X.}~\bibnamefont
  {Wu}}, \bibinfo {author} {\bibfnamefont {Y.}~\bibnamefont {Hu}}, \bibinfo
  {author} {\bibfnamefont {M.}~\bibnamefont {Ruan}}, \bibinfo {author}
  {\bibfnamefont {N.~K.}\ \bibnamefont {Madiomanana}}, \bibinfo {author}
  {\bibfnamefont {J.}~\bibnamefont {Hankinson}}, \bibinfo {author}
  {\bibfnamefont {M.}~\bibnamefont {Sprinkle}}, \bibinfo {author}
  {\bibfnamefont {C.}~\bibnamefont {Berger}},\ and\ \bibinfo {author}
  {\bibfnamefont {W.~A.}\ \bibnamefont {de~Heer}},\ }\href
  {https://doi.org/10.1063/1.3266524} {\bibfield  {journal} {\bibinfo
  {journal} {Applied Physics Letters}\ }\textbf {\bibinfo {volume} {95}},\
  \bibinfo {pages} {223108} (\bibinfo {year} {2009})}\BibitemShut {NoStop}%
\bibitem [{\citenamefont {Fujita}\ and\ \citenamefont
  {Suzuki}(2016)}]{fujita2016}%
  \BibitemOpen
  \bibfield  {author} {\bibinfo {author} {\bibfnamefont {S.}~\bibnamefont
  {Fujita}}\ and\ \bibinfo {author} {\bibfnamefont {A.}~\bibnamefont
  {Suzuki}},\ }\href {https://doi.org/10.1007/s10773-016-3106-8} {\bibfield
  {journal} {\bibinfo  {journal} {International Journal of Theoretical
  Physics}\ }\textbf {\bibinfo {volume} {55}} (\bibinfo {year}
  {2016})}\BibitemShut {NoStop}%
\bibitem [{\citenamefont {Jiang}\ \emph {et~al.}(2007)\citenamefont {Jiang},
  \citenamefont {Zhang}, \citenamefont {Tan}, \citenamefont {Stormer},\ and\
  \citenamefont {Kim}}]{jiang2007}%
  \BibitemOpen
  \bibfield  {author} {\bibinfo {author} {\bibfnamefont {Z.}~\bibnamefont
  {Jiang}}, \bibinfo {author} {\bibfnamefont {Y.}~\bibnamefont {Zhang}},
  \bibinfo {author} {\bibfnamefont {Y.-W.}\ \bibnamefont {Tan}}, \bibinfo
  {author} {\bibfnamefont {H.}~\bibnamefont {Stormer}},\ and\ \bibinfo {author}
  {\bibfnamefont {P.}~\bibnamefont {Kim}},\ }\href
  {https://doi.org/https://doi.org/10.1016/j.ssc.2007.02.046} {\bibfield
  {journal} {\bibinfo  {journal} {Solid State Communications}\ }\textbf
  {\bibinfo {volume} {143}},\ \bibinfo {pages} {14} (\bibinfo {year}
  {2007})}\BibitemShut {NoStop}%
\bibitem [{\citenamefont {Gusynin}\ \emph {et~al.}(2008)\citenamefont
  {Gusynin}, \citenamefont {Miransky}, \citenamefont {Sharapov},\ and\
  \citenamefont {Shovkovy}}]{gusynin2008}%
  \BibitemOpen
  \bibfield  {author} {\bibinfo {author} {\bibfnamefont {V.~P.}\ \bibnamefont
  {Gusynin}}, \bibinfo {author} {\bibfnamefont {V.~A.}\ \bibnamefont
  {Miransky}}, \bibinfo {author} {\bibfnamefont {S.~G.}\ \bibnamefont
  {Sharapov}},\ and\ \bibinfo {author} {\bibfnamefont {I.~A.}\ \bibnamefont
  {Shovkovy}},\ }\href {https://doi.org/10.1063/1.2981387} {\bibfield
  {journal} {\bibinfo  {journal} {Low Temperature Physics}\ }\textbf {\bibinfo
  {volume} {34}},\ \bibinfo {pages} {778} (\bibinfo {year} {2008})}\BibitemShut
  {NoStop}%
\bibitem [{\citenamefont {Lado}\ \emph {et~al.}(2013)\citenamefont {Lado},
  \citenamefont {Gonz\'alez},\ and\ \citenamefont
  {Fern\'andez-Rossier}}]{lado2013}%
  \BibitemOpen
  \bibfield  {author} {\bibinfo {author} {\bibfnamefont {J.~L.}\ \bibnamefont
  {Lado}}, \bibinfo {author} {\bibfnamefont {J.~W.}\ \bibnamefont
  {Gonz\'alez}},\ and\ \bibinfo {author} {\bibfnamefont {J.}~\bibnamefont
  {Fern\'andez-Rossier}},\ }\href {https://doi.org/10.1103/PhysRevB.88.035448}
  {\bibfield  {journal} {\bibinfo  {journal} {Phys. Rev. B}\ }\textbf {\bibinfo
  {volume} {88}},\ \bibinfo {pages} {035448} (\bibinfo {year}
  {2013})}\BibitemShut {NoStop}%
\bibitem [{\citenamefont {Amet}\ \emph {et~al.}(2014)\citenamefont {Amet},
  \citenamefont {Williams}, \citenamefont {Watanabe}, \citenamefont
  {Taniguchi},\ and\ \citenamefont {Goldhaber-Gordon}}]{amet2014}%
  \BibitemOpen
  \bibfield  {author} {\bibinfo {author} {\bibfnamefont {F.}~\bibnamefont
  {Amet}}, \bibinfo {author} {\bibfnamefont {J.~R.}\ \bibnamefont {Williams}},
  \bibinfo {author} {\bibfnamefont {K.}~\bibnamefont {Watanabe}}, \bibinfo
  {author} {\bibfnamefont {T.}~\bibnamefont {Taniguchi}},\ and\ \bibinfo
  {author} {\bibfnamefont {D.}~\bibnamefont {Goldhaber-Gordon}},\ }\href@noop
  {} {\bibfield  {journal} {\bibinfo  {journal} {Phys. Rev. Lett.}\ }\textbf
  {\bibinfo {volume} {112}},\ \bibinfo {pages} {196601} (\bibinfo {year}
  {2014})}\BibitemShut {NoStop}%
\bibitem [{\citenamefont {Zimmermann}\ \emph {et~al.}(2017)\citenamefont
  {Zimmermann}, \citenamefont {Jordan}, \citenamefont {Gay}, \citenamefont
  {Watanabe}, \citenamefont {Taniguchi}, \citenamefont {Han}, \citenamefont
  {Bouchiat}, \citenamefont {Sellier},\ and\ \citenamefont
  {Sac{\'{e}}p{\'{e}}}}]{zimmermann2017}%
  \BibitemOpen
  \bibfield  {author} {\bibinfo {author} {\bibfnamefont {K.}~\bibnamefont
  {Zimmermann}}, \bibinfo {author} {\bibfnamefont {A.}~\bibnamefont {Jordan}},
  \bibinfo {author} {\bibfnamefont {F.}~\bibnamefont {Gay}}, \bibinfo {author}
  {\bibfnamefont {K.}~\bibnamefont {Watanabe}}, \bibinfo {author}
  {\bibfnamefont {T.}~\bibnamefont {Taniguchi}}, \bibinfo {author}
  {\bibfnamefont {Z.}~\bibnamefont {Han}}, \bibinfo {author} {\bibfnamefont
  {V.}~\bibnamefont {Bouchiat}}, \bibinfo {author} {\bibfnamefont
  {H.}~\bibnamefont {Sellier}},\ and\ \bibinfo {author} {\bibfnamefont
  {B.}~\bibnamefont {Sac{\'{e}}p{\'{e}}}},\ }\href@noop {} {\bibfield
  {journal} {\bibinfo  {journal} {Nature Communications}\ }\textbf {\bibinfo
  {volume} {8}} (\bibinfo {year} {2017})}\BibitemShut {NoStop}%
\bibitem [{\citenamefont {Ding}\ \emph {et~al.}(2018)\citenamefont {Ding},
  \citenamefont {Lim}, \citenamefont {Su},\ and\ \citenamefont
  {Weng}}]{ding2018}%
  \BibitemOpen
  \bibfield  {author} {\bibinfo {author} {\bibfnamefont {K.-H.}\ \bibnamefont
  {Ding}}, \bibinfo {author} {\bibfnamefont {L.-K.}\ \bibnamefont {Lim}},
  \bibinfo {author} {\bibfnamefont {G.}~\bibnamefont {Su}},\ and\ \bibinfo
  {author} {\bibfnamefont {Z.-Y.}\ \bibnamefont {Weng}},\ }\href
  {https://doi.org/10.1103/PhysRevB.97.035123} {\bibfield  {journal} {\bibinfo
  {journal} {Phys. Rev. B}\ }\textbf {\bibinfo {volume} {97}},\ \bibinfo
  {pages} {035123} (\bibinfo {year} {2018})}\BibitemShut {NoStop}%
\bibitem [{\citenamefont {Liu}\ \emph {et~al.}(2022)\citenamefont {Liu},
  \citenamefont {Farahi}, \citenamefont {Chiu}, \citenamefont {Papic},
  \citenamefont {Watanabe}, \citenamefont {Taniguchi}, \citenamefont
  {Zaletel},\ and\ \citenamefont {Yazdani}}]{liu2022}%
  \BibitemOpen
  \bibfield  {author} {\bibinfo {author} {\bibfnamefont {X.}~\bibnamefont
  {Liu}}, \bibinfo {author} {\bibfnamefont {G.}~\bibnamefont {Farahi}},
  \bibinfo {author} {\bibfnamefont {C.-L.}\ \bibnamefont {Chiu}}, \bibinfo
  {author} {\bibfnamefont {Z.}~\bibnamefont {Papic}}, \bibinfo {author}
  {\bibfnamefont {K.}~\bibnamefont {Watanabe}}, \bibinfo {author}
  {\bibfnamefont {T.}~\bibnamefont {Taniguchi}}, \bibinfo {author}
  {\bibfnamefont {M.~P.}\ \bibnamefont {Zaletel}},\ and\ \bibinfo {author}
  {\bibfnamefont {A.}~\bibnamefont {Yazdani}},\ }\href
  {https://doi.org/10.1126/science.abm3770} {\bibfield  {journal} {\bibinfo
  {journal} {Science}\ }\textbf {\bibinfo {volume} {375}},\ \bibinfo {pages}
  {321} (\bibinfo {year} {2022})}\BibitemShut {NoStop}%
\bibitem [{\citenamefont {Wei}\ \emph {et~al.}(2018)\citenamefont {Wei},
  \citenamefont {van~der Sar}, \citenamefont {Lee}, \citenamefont {Watanabe},
  \citenamefont {Taniguchi}, \citenamefont {Halperin},\ and\ \citenamefont
  {Yacoby}}]{wei2018}%
  \BibitemOpen
  \bibfield  {author} {\bibinfo {author} {\bibfnamefont {D.~S.}\ \bibnamefont
  {Wei}}, \bibinfo {author} {\bibfnamefont {T.}~\bibnamefont {van~der Sar}},
  \bibinfo {author} {\bibfnamefont {S.~H.}\ \bibnamefont {Lee}}, \bibinfo
  {author} {\bibfnamefont {K.}~\bibnamefont {Watanabe}}, \bibinfo {author}
  {\bibfnamefont {T.}~\bibnamefont {Taniguchi}}, \bibinfo {author}
  {\bibfnamefont {B.~I.}\ \bibnamefont {Halperin}},\ and\ \bibinfo {author}
  {\bibfnamefont {A.}~\bibnamefont {Yacoby}},\ }\href@noop {} {\bibfield
  {journal} {\bibinfo  {journal} {Science}\ }\textbf {\bibinfo {volume}
  {362}},\ \bibinfo {pages} {229} (\bibinfo {year} {2018})}\BibitemShut
  {NoStop}%
\bibitem [{\citenamefont {Assouline}\ \emph {et~al.}(2021)\citenamefont
  {Assouline}, \citenamefont {Jo}, \citenamefont {Brasseur}, \citenamefont
  {Watanabe}, \citenamefont {Taniguchi}, \citenamefont {Jolicoeur},
  \citenamefont {Glattli}, \citenamefont {Kumada}, \citenamefont {Roche},
  \citenamefont {Parmentier},\ and\ \citenamefont {Roulleau}}]{assouline2021}%
  \BibitemOpen
  \bibfield  {author} {\bibinfo {author} {\bibfnamefont {A.}~\bibnamefont
  {Assouline}}, \bibinfo {author} {\bibfnamefont {M.}~\bibnamefont {Jo}},
  \bibinfo {author} {\bibfnamefont {P.}~\bibnamefont {Brasseur}}, \bibinfo
  {author} {\bibfnamefont {K.}~\bibnamefont {Watanabe}}, \bibinfo {author}
  {\bibfnamefont {T.}~\bibnamefont {Taniguchi}}, \bibinfo {author}
  {\bibfnamefont {T.}~\bibnamefont {Jolicoeur}}, \bibinfo {author}
  {\bibfnamefont {D.}~\bibnamefont {Glattli}}, \bibinfo {author} {\bibfnamefont
  {N.}~\bibnamefont {Kumada}}, \bibinfo {author} {\bibfnamefont
  {P.}~\bibnamefont {Roche}}, \bibinfo {author} {\bibfnamefont
  {F.}~\bibnamefont {Parmentier}},\ and\ \bibinfo {author} {\bibfnamefont
  {P.}~\bibnamefont {Roulleau}},\ }\href
  {https://doi.org/10.1038/s41567-021-01411-z} {\bibfield  {journal} {\bibinfo
  {journal} {Nature Physics}\ }\textbf {\bibinfo {volume} {17}},\ \bibinfo
  {pages} {1} (\bibinfo {year} {2021})}\BibitemShut {NoStop}%
\bibitem [{\citenamefont {Pierce}\ \emph {et~al.}(2021)\citenamefont {Pierce},
  \citenamefont {Xie}, \citenamefont {Lee}, \citenamefont {Forrester},
  \citenamefont {Wei}, \citenamefont {Watanabe}, \citenamefont {Taniguchi},
  \citenamefont {Halperin},\ and\ \citenamefont {Yacoby}}]{pierce2021}%
  \BibitemOpen
  \bibfield  {author} {\bibinfo {author} {\bibfnamefont {A.~T.}\ \bibnamefont
  {Pierce}}, \bibinfo {author} {\bibfnamefont {Y.}~\bibnamefont {Xie}},
  \bibinfo {author} {\bibfnamefont {S.~H.}\ \bibnamefont {Lee}}, \bibinfo
  {author} {\bibfnamefont {P.~R.}\ \bibnamefont {Forrester}}, \bibinfo {author}
  {\bibfnamefont {D.~S.}\ \bibnamefont {Wei}}, \bibinfo {author} {\bibfnamefont
  {K.}~\bibnamefont {Watanabe}}, \bibinfo {author} {\bibfnamefont
  {T.}~\bibnamefont {Taniguchi}}, \bibinfo {author} {\bibfnamefont {B.~I.}\
  \bibnamefont {Halperin}},\ and\ \bibinfo {author} {\bibfnamefont
  {A.}~\bibnamefont {Yacoby}},\ }\href@noop {} {\bibfield  {journal} {\bibinfo
  {journal} {Nature Physics}\ }\textbf {\bibinfo {volume} {18}},\ \bibinfo
  {pages} {37} (\bibinfo {year} {2021})}\BibitemShut {NoStop}%
\bibitem [{\citenamefont {Lian}\ and\ \citenamefont
  {Goerbig}(2017)}]{lian2017}%
  \BibitemOpen
  \bibfield  {author} {\bibinfo {author} {\bibfnamefont {Y.}~\bibnamefont
  {Lian}}\ and\ \bibinfo {author} {\bibfnamefont {M.~O.}\ \bibnamefont
  {Goerbig}},\ }\href {https://doi.org/10.1103/PhysRevB.95.245428} {\bibfield
  {journal} {\bibinfo  {journal} {Phys. Rev. B}\ }\textbf {\bibinfo {volume}
  {95}},\ \bibinfo {pages} {245428} (\bibinfo {year} {2017})}\BibitemShut
  {NoStop}%
\bibitem [{\citenamefont {Berman}\ \emph {et~al.}(2022)\citenamefont {Berman},
  \citenamefont {Kezerashvili}, \citenamefont {Lozovik},\ and\ \citenamefont
  {Ziegler}}]{berman2022}%
  \BibitemOpen
  \bibfield  {author} {\bibinfo {author} {\bibfnamefont {O.~L.}\ \bibnamefont
  {Berman}}, \bibinfo {author} {\bibfnamefont {R.~Y.}\ \bibnamefont
  {Kezerashvili}}, \bibinfo {author} {\bibfnamefont {Y.~E.}\ \bibnamefont
  {Lozovik}},\ and\ \bibinfo {author} {\bibfnamefont {K.~G.}\ \bibnamefont
  {Ziegler}},\ }\href@noop {} {\bibfield  {journal} {\bibinfo  {journal}
  {Scientific Reports}\ }\textbf {\bibinfo {volume} {12}} (\bibinfo {year}
  {2022})}\BibitemShut {NoStop}%
\bibitem [{\citenamefont {Wei}\ \emph {et~al.}(2017)\citenamefont {Wei},
  \citenamefont {van~der Sar}, \citenamefont {Sanchez-Yamagishi}, \citenamefont
  {Watanabe}, \citenamefont {Taniguchi}, \citenamefont {Jarillo-Herrero},
  \citenamefont {Halperin},\ and\ \citenamefont {Yacoby}}]{wei2017}%
  \BibitemOpen
  \bibfield  {author} {\bibinfo {author} {\bibfnamefont {D.~S.}\ \bibnamefont
  {Wei}}, \bibinfo {author} {\bibfnamefont {T.}~\bibnamefont {van~der Sar}},
  \bibinfo {author} {\bibfnamefont {J.~D.}\ \bibnamefont {Sanchez-Yamagishi}},
  \bibinfo {author} {\bibfnamefont {K.}~\bibnamefont {Watanabe}}, \bibinfo
  {author} {\bibfnamefont {T.}~\bibnamefont {Taniguchi}}, \bibinfo {author}
  {\bibfnamefont {P.}~\bibnamefont {Jarillo-Herrero}}, \bibinfo {author}
  {\bibfnamefont {B.~I.}\ \bibnamefont {Halperin}},\ and\ \bibinfo {author}
  {\bibfnamefont {A.}~\bibnamefont {Yacoby}},\ }\href@noop {} {\bibfield
  {journal} {\bibinfo  {journal} {Science Advances}\ }\textbf {\bibinfo
  {volume} {3}} (\bibinfo {year} {2017})}\BibitemShut {NoStop}%
\bibitem [{\citenamefont {Ara{\'{u}}jo}\ \emph {et~al.}(2020)\citenamefont
  {Ara{\'{u}}jo}, \citenamefont {da~Costa}, \citenamefont {Nascimento},\ and\
  \citenamefont {Pereira}}]{araujo2020}%
  \BibitemOpen
  \bibfield  {author} {\bibinfo {author} {\bibfnamefont {F.~R.~V.}\
  \bibnamefont {Ara{\'{u}}jo}}, \bibinfo {author} {\bibfnamefont {D.~R.}\
  \bibnamefont {da~Costa}}, \bibinfo {author} {\bibfnamefont {A.~C.~S.}\
  \bibnamefont {Nascimento}},\ and\ \bibinfo {author} {\bibfnamefont {J.~M.}\
  \bibnamefont {Pereira}},\ }\href@noop {} {\bibfield  {journal} {\bibinfo
  {journal} {Journal of Physics: Condensed Matter}\ }\textbf {\bibinfo {volume}
  {32}},\ \bibinfo {pages} {425501} (\bibinfo {year} {2020})}\BibitemShut
  {NoStop}%
\bibitem [{\citenamefont {Carrega}\ \emph {et~al.}(2021)\citenamefont
  {Carrega}, \citenamefont {Chirolli}, \citenamefont {Heun},\ and\
  \citenamefont {Sorba}}]{carrega2021}%
  \BibitemOpen
  \bibfield  {author} {\bibinfo {author} {\bibfnamefont {M.}~\bibnamefont
  {Carrega}}, \bibinfo {author} {\bibfnamefont {L.}~\bibnamefont {Chirolli}},
  \bibinfo {author} {\bibfnamefont {S.}~\bibnamefont {Heun}},\ and\ \bibinfo
  {author} {\bibfnamefont {L.}~\bibnamefont {Sorba}},\ }\href@noop {}
  {\bibfield  {journal} {\bibinfo  {journal} {Nature Reviews Physics}\ }\textbf
  {\bibinfo {volume} {3}},\ \bibinfo {pages} {698} (\bibinfo {year}
  {2021})}\BibitemShut {NoStop}%
\bibitem [{\citenamefont {Jo}\ \emph {et~al.}(2021)\citenamefont {Jo},
  \citenamefont {Brasseur}, \citenamefont {Assouline}, \citenamefont {Fleury},
  \citenamefont {Sim}, \citenamefont {Watanabe}, \citenamefont {Taniguchi},
  \citenamefont {Dumnernpanich}, \citenamefont {Roche}, \citenamefont
  {Glattli}, \citenamefont {Kumada}, \citenamefont {Parmentier},\ and\
  \citenamefont {Roulleau}}]{jo2021}%
  \BibitemOpen
  \bibfield  {author} {\bibinfo {author} {\bibfnamefont {M.}~\bibnamefont
  {Jo}}, \bibinfo {author} {\bibfnamefont {P.}~\bibnamefont {Brasseur}},
  \bibinfo {author} {\bibfnamefont {A.}~\bibnamefont {Assouline}}, \bibinfo
  {author} {\bibfnamefont {G.}~\bibnamefont {Fleury}}, \bibinfo {author}
  {\bibfnamefont {H.-S.}\ \bibnamefont {Sim}}, \bibinfo {author} {\bibfnamefont
  {K.}~\bibnamefont {Watanabe}}, \bibinfo {author} {\bibfnamefont
  {T.}~\bibnamefont {Taniguchi}}, \bibinfo {author} {\bibfnamefont
  {W.}~\bibnamefont {Dumnernpanich}}, \bibinfo {author} {\bibfnamefont
  {P.}~\bibnamefont {Roche}}, \bibinfo {author} {\bibfnamefont {D.~C.}\
  \bibnamefont {Glattli}}, \bibinfo {author} {\bibfnamefont {N.}~\bibnamefont
  {Kumada}}, \bibinfo {author} {\bibfnamefont {F.~D.}\ \bibnamefont
  {Parmentier}},\ and\ \bibinfo {author} {\bibfnamefont {P.}~\bibnamefont
  {Roulleau}},\ }\href {https://doi.org/10.1103/PhysRevLett.126.146803}
  {\bibfield  {journal} {\bibinfo  {journal} {Phys. Rev. Lett.}\ }\textbf
  {\bibinfo {volume} {126}},\ \bibinfo {pages} {146803} (\bibinfo {year}
  {2021})}\BibitemShut {NoStop}%
\bibitem [{\citenamefont {D{\'{e}}prez}\ \emph {et~al.}(2021)\citenamefont
  {D{\'{e}}prez}, \citenamefont {Veyrat}, \citenamefont {Vignaud},
  \citenamefont {Nayak}, \citenamefont {Watanabe}, \citenamefont {Taniguchi},
  \citenamefont {Gay}, \citenamefont {Sellier},\ and\ \citenamefont
  {Sac{\'{e}}p{\'{e}}}}]{deprez2021}%
  \BibitemOpen
  \bibfield  {author} {\bibinfo {author} {\bibfnamefont {C.}~\bibnamefont
  {D{\'{e}}prez}}, \bibinfo {author} {\bibfnamefont {L.}~\bibnamefont
  {Veyrat}}, \bibinfo {author} {\bibfnamefont {H.}~\bibnamefont {Vignaud}},
  \bibinfo {author} {\bibfnamefont {G.}~\bibnamefont {Nayak}}, \bibinfo
  {author} {\bibfnamefont {K.}~\bibnamefont {Watanabe}}, \bibinfo {author}
  {\bibfnamefont {T.}~\bibnamefont {Taniguchi}}, \bibinfo {author}
  {\bibfnamefont {F.}~\bibnamefont {Gay}}, \bibinfo {author} {\bibfnamefont
  {H.}~\bibnamefont {Sellier}},\ and\ \bibinfo {author} {\bibfnamefont
  {B.}~\bibnamefont {Sac{\'{e}}p{\'{e}}}},\ }\href@noop {} {\bibfield
  {journal} {\bibinfo  {journal} {Nature Nanotechnology}\ }\textbf {\bibinfo
  {volume} {16}},\ \bibinfo {pages} {555} (\bibinfo {year} {2021})}\BibitemShut
  {NoStop}%
\bibitem [{\citenamefont {Johnson}\ \emph {et~al.}(2021)\citenamefont
  {Johnson}, \citenamefont {Turner},\ and\ \citenamefont
  {McMorran}}]{johnson2021}%
  \BibitemOpen
  \bibfield  {author} {\bibinfo {author} {\bibfnamefont {C.~W.}\ \bibnamefont
  {Johnson}}, \bibinfo {author} {\bibfnamefont {A.~E.}\ \bibnamefont
  {Turner}},\ and\ \bibinfo {author} {\bibfnamefont {B.~J.}\ \bibnamefont
  {McMorran}},\ }\href {https://doi.org/10.1103/PhysRevResearch.3.043009}
  {\bibfield  {journal} {\bibinfo  {journal} {Phys. Rev. Research}\ }\textbf
  {\bibinfo {volume} {3}},\ \bibinfo {pages} {043009} (\bibinfo {year}
  {2021})}\BibitemShut {NoStop}%
\bibitem [{\citenamefont {Marconcini}\ and\ \citenamefont
  {Macucci}(2022)}]{marconcini2022}%
  \BibitemOpen
  \bibfield  {author} {\bibinfo {author} {\bibfnamefont {P.}~\bibnamefont
  {Marconcini}}\ and\ \bibinfo {author} {\bibfnamefont {M.}~\bibnamefont
  {Macucci}},\ }\bibfield  {journal} {\bibinfo  {journal} {Nanomaterials}\
  }\textbf {\bibinfo {volume} {12}},\ \href
  {https://doi.org/10.3390/nano12071087} {10.3390/nano12071087} (\bibinfo
  {year} {2022})\BibitemShut {NoStop}%
\bibitem [{\citenamefont {Mirzakhani}\ \emph {et~al.}(2022)\citenamefont
  {Mirzakhani}, \citenamefont {Myoung}, \citenamefont {Peeters},\ and\
  \citenamefont {Park}}]{mirzakhani2022}%
  \BibitemOpen
  \bibfield  {author} {\bibinfo {author} {\bibfnamefont {M.}~\bibnamefont
  {Mirzakhani}}, \bibinfo {author} {\bibfnamefont {N.}~\bibnamefont {Myoung}},
  \bibinfo {author} {\bibfnamefont {F.~M.}\ \bibnamefont {Peeters}},\ and\
  \bibinfo {author} {\bibfnamefont {H.~C.}\ \bibnamefont {Park}},\ }\href@noop
  {} {} (\bibinfo {year} {2022})\BibitemShut {NoStop}%
\bibitem [{\citenamefont {Edlbauer}\ \emph {et~al.}(2022)\citenamefont
  {Edlbauer}, \citenamefont {Wang}, \citenamefont {Crozes}, \citenamefont
  {Perrier}, \citenamefont {Ouacel}, \citenamefont {Geffroy}, \citenamefont
  {Georgiou}, \citenamefont {Chatzikyriakou}, \citenamefont {Lacerda-Santos},
  \citenamefont {Waintal}, \citenamefont {Glattli}, \citenamefont {Roulleau},
  \citenamefont {Nath}, \citenamefont {Kataoka}, \citenamefont
  {Splettstoesser}, \citenamefont {Acciai}, \citenamefont {da~Silva~Figueira},
  \citenamefont {Öztas}, \citenamefont {Trellakis}, \citenamefont {Grange},
  \citenamefont {Yevtushenko}, \citenamefont {Birner},\ and\ \citenamefont
  {Bäuerle}}]{edlbauer2022}%
  \BibitemOpen
  \bibfield  {author} {\bibinfo {author} {\bibfnamefont {H.}~\bibnamefont
  {Edlbauer}}, \bibinfo {author} {\bibfnamefont {J.}~\bibnamefont {Wang}},
  \bibinfo {author} {\bibfnamefont {T.}~\bibnamefont {Crozes}}, \bibinfo
  {author} {\bibfnamefont {P.}~\bibnamefont {Perrier}}, \bibinfo {author}
  {\bibfnamefont {S.}~\bibnamefont {Ouacel}}, \bibinfo {author} {\bibfnamefont
  {C.}~\bibnamefont {Geffroy}}, \bibinfo {author} {\bibfnamefont
  {G.}~\bibnamefont {Georgiou}}, \bibinfo {author} {\bibfnamefont
  {E.}~\bibnamefont {Chatzikyriakou}}, \bibinfo {author} {\bibfnamefont
  {A.}~\bibnamefont {Lacerda-Santos}}, \bibinfo {author} {\bibfnamefont
  {X.}~\bibnamefont {Waintal}}, \bibinfo {author} {\bibfnamefont {D.~C.}\
  \bibnamefont {Glattli}}, \bibinfo {author} {\bibfnamefont {P.}~\bibnamefont
  {Roulleau}}, \bibinfo {author} {\bibfnamefont {J.}~\bibnamefont {Nath}},
  \bibinfo {author} {\bibfnamefont {M.}~\bibnamefont {Kataoka}}, \bibinfo
  {author} {\bibfnamefont {J.}~\bibnamefont {Splettstoesser}}, \bibinfo
  {author} {\bibfnamefont {M.}~\bibnamefont {Acciai}}, \bibinfo {author}
  {\bibfnamefont {M.~C.}\ \bibnamefont {da~Silva~Figueira}}, \bibinfo {author}
  {\bibfnamefont {K.}~\bibnamefont {Öztas}}, \bibinfo {author} {\bibfnamefont
  {A.}~\bibnamefont {Trellakis}}, \bibinfo {author} {\bibfnamefont
  {T.}~\bibnamefont {Grange}}, \bibinfo {author} {\bibfnamefont {O.~M.}\
  \bibnamefont {Yevtushenko}}, \bibinfo {author} {\bibfnamefont
  {S.}~\bibnamefont {Birner}},\ and\ \bibinfo {author} {\bibfnamefont
  {C.}~\bibnamefont {Bäuerle}},\ }\href@noop {} {\bibfield  {journal}
  {\bibinfo  {journal} {{EPJ} Quantum Technology}\ }\textbf {\bibinfo {volume}
  {9}} (\bibinfo {year} {2022})}\BibitemShut {NoStop}%
\bibitem [{\citenamefont {Furtado}\ \emph {et~al.}(2022)\citenamefont
  {Furtado}, \citenamefont {Ramos}, \citenamefont {Silva}, \citenamefont
  {Bachelard},\ and\ \citenamefont {Santos}}]{furtado2022}%
  \BibitemOpen
  \bibfield  {author} {\bibinfo {author} {\bibfnamefont {J.}~\bibnamefont
  {Furtado}}, \bibinfo {author} {\bibfnamefont {A.~C.~A.}\ \bibnamefont
  {Ramos}}, \bibinfo {author} {\bibfnamefont {J.~E.~G.}\ \bibnamefont {Silva}},
  \bibinfo {author} {\bibfnamefont {R.}~\bibnamefont {Bachelard}},\ and\
  \bibinfo {author} {\bibfnamefont {A.~C.}\ \bibnamefont {Santos}},\
  }\href@noop {} {} (\bibinfo {year} {2022})\BibitemShut {NoStop}%
\bibitem [{\citenamefont {Dean}\ \emph {et~al.}(2010)\citenamefont {Dean},
  \citenamefont {Young}, \citenamefont {Meric}, \citenamefont {Lee},
  \citenamefont {Wang}, \citenamefont {Sorgenfrei}, \citenamefont {Watanabe},
  \citenamefont {Taniguchi}, \citenamefont {Kim}, \citenamefont {Shepard},\
  and\ \citenamefont {Hone}}]{dean2010}%
  \BibitemOpen
  \bibfield  {author} {\bibinfo {author} {\bibfnamefont {C.~R.}\ \bibnamefont
  {Dean}}, \bibinfo {author} {\bibfnamefont {A.~F.}\ \bibnamefont {Young}},
  \bibinfo {author} {\bibfnamefont {I.}~\bibnamefont {Meric}}, \bibinfo
  {author} {\bibfnamefont {C.}~\bibnamefont {Lee}}, \bibinfo {author}
  {\bibfnamefont {L.}~\bibnamefont {Wang}}, \bibinfo {author} {\bibfnamefont
  {S.}~\bibnamefont {Sorgenfrei}}, \bibinfo {author} {\bibfnamefont
  {K.}~\bibnamefont {Watanabe}}, \bibinfo {author} {\bibfnamefont
  {T.}~\bibnamefont {Taniguchi}}, \bibinfo {author} {\bibfnamefont
  {P.}~\bibnamefont {Kim}}, \bibinfo {author} {\bibfnamefont {K.~L.}\
  \bibnamefont {Shepard}},\ and\ \bibinfo {author} {\bibfnamefont
  {J.}~\bibnamefont {Hone}},\ }\href@noop {} {\bibfield  {journal} {\bibinfo
  {journal} {Nature Nanotechnology}\ }\textbf {\bibinfo {volume} {5}},\
  \bibinfo {pages} {722} (\bibinfo {year} {2010})}\BibitemShut {NoStop}%
\bibitem [{\citenamefont {Wang}\ \emph {et~al.}(2017)\citenamefont {Wang},
  \citenamefont {Ma},\ and\ \citenamefont {Sun}}]{wang2017}%
  \BibitemOpen
  \bibfield  {author} {\bibinfo {author} {\bibfnamefont {J.}~\bibnamefont
  {Wang}}, \bibinfo {author} {\bibfnamefont {F.}~\bibnamefont {Ma}},\ and\
  \bibinfo {author} {\bibfnamefont {M.}~\bibnamefont {Sun}},\ }\href
  {https://doi.org/10.1039/C7RA00260B} {\bibfield  {journal} {\bibinfo
  {journal} {RSC Advances}\ }\textbf {\bibinfo {volume} {7}} (\bibinfo {year}
  {2017})}\BibitemShut {NoStop}%
\bibitem [{\citenamefont {Oroszlány}\ \emph {et~al.}(2008)\citenamefont
  {Oroszlány}, \citenamefont {Rakyta}, \citenamefont {Kormányos},
  \citenamefont {Lambert},\ and\ \citenamefont {Cserti}}]{oroszlany2008}%
  \BibitemOpen
  \bibfield  {author} {\bibinfo {author} {\bibfnamefont {L.}~\bibnamefont
  {Oroszlány}}, \bibinfo {author} {\bibfnamefont {P.}~\bibnamefont {Rakyta}},
  \bibinfo {author} {\bibfnamefont {A.}~\bibnamefont {Kormányos}}, \bibinfo
  {author} {\bibfnamefont {C.}~\bibnamefont {Lambert}},\ and\ \bibinfo {author}
  {\bibfnamefont {J.}~\bibnamefont {Cserti}},\ }\href
  {https://doi.org/10.1103/PhysRevB.77.081403} {\bibfield  {journal} {\bibinfo
  {journal} {Physical Review B}\ }\textbf {\bibinfo {volume} {77}},\ \bibinfo
  {pages} {081403} (\bibinfo {year} {2008})}\BibitemShut {NoStop}%
\bibitem [{\citenamefont {Williams}\ and\ \citenamefont
  {Marcus}(2011)}]{williams2011}%
  \BibitemOpen
  \bibfield  {author} {\bibinfo {author} {\bibfnamefont {J.~R.}\ \bibnamefont
  {Williams}}\ and\ \bibinfo {author} {\bibfnamefont {C.~M.}\ \bibnamefont
  {Marcus}},\ }\href {https://doi.org/10.1103/PhysRevLett.107.046602}
  {\bibfield  {journal} {\bibinfo  {journal} {Phys. Rev. Lett.}\ }\textbf
  {\bibinfo {volume} {107}},\ \bibinfo {pages} {046602} (\bibinfo {year}
  {2011})}\BibitemShut {NoStop}%
\bibitem [{\citenamefont {Carmier}\ \emph {et~al.}(2011)\citenamefont
  {Carmier}, \citenamefont {Lewenkopf},\ and\ \citenamefont
  {Ullmo}}]{carmier2011}%
  \BibitemOpen
  \bibfield  {author} {\bibinfo {author} {\bibfnamefont {P.}~\bibnamefont
  {Carmier}}, \bibinfo {author} {\bibfnamefont {C.}~\bibnamefont {Lewenkopf}},\
  and\ \bibinfo {author} {\bibfnamefont {D.}~\bibnamefont {Ullmo}},\
  }\href@noop {} {\bibfield  {journal} {\bibinfo  {journal} {Physical Review
  B}\ }\textbf {\bibinfo {volume} {84}} (\bibinfo {year} {2011})}\BibitemShut
  {NoStop}%
\bibitem [{\citenamefont {Milovanović}\ \emph {et~al.}(2014)\citenamefont
  {Milovanović}, \citenamefont {Ramezani~Masir},\ and\ \citenamefont
  {Peeters}}]{milovanovic2014}%
  \BibitemOpen
  \bibfield  {author} {\bibinfo {author} {\bibfnamefont {S.~P.}\ \bibnamefont
  {Milovanović}}, \bibinfo {author} {\bibfnamefont {M.}~\bibnamefont
  {Ramezani~Masir}},\ and\ \bibinfo {author} {\bibfnamefont {F.~M.}\
  \bibnamefont {Peeters}},\ }\href {https://doi.org/10.1063/1.4896769}
  {\bibfield  {journal} {\bibinfo  {journal} {Applied Physics Letters}\
  }\textbf {\bibinfo {volume} {105}},\ \bibinfo {pages} {123507} (\bibinfo
  {year} {2014})}\BibitemShut {NoStop}%
\bibitem [{\citenamefont {Taychatanapat}\ \emph {et~al.}(2015)\citenamefont
  {Taychatanapat}, \citenamefont {Tan}, \citenamefont {Yeo}, \citenamefont
  {Watanabe}, \citenamefont {Taniguchi},\ and\ \citenamefont
  {Özyilmaz}}]{taychatanapat2015}%
  \BibitemOpen
  \bibfield  {author} {\bibinfo {author} {\bibfnamefont {T.}~\bibnamefont
  {Taychatanapat}}, \bibinfo {author} {\bibfnamefont {J.~Y.}\ \bibnamefont
  {Tan}}, \bibinfo {author} {\bibfnamefont {Y.}~\bibnamefont {Yeo}}, \bibinfo
  {author} {\bibfnamefont {K.}~\bibnamefont {Watanabe}}, \bibinfo {author}
  {\bibfnamefont {T.}~\bibnamefont {Taniguchi}},\ and\ \bibinfo {author}
  {\bibfnamefont {B.}~\bibnamefont {Özyilmaz}},\ }\href@noop {} {\bibfield
  {journal} {\bibinfo  {journal} {Nature Communications}\ }\textbf {\bibinfo
  {volume} {6}} (\bibinfo {year} {2015})}\BibitemShut {NoStop}%
\bibitem [{\citenamefont {Rickhaus}\ \emph {et~al.}(2015)\citenamefont
  {Rickhaus}, \citenamefont {Makk}, \citenamefont {Liu}, \citenamefont
  {Tóvári}, \citenamefont {Weiss}, \citenamefont {Maurand}, \citenamefont
  {Richter},\ and\ \citenamefont {Schoenenberger}}]{rickhaus2015_2}%
  \BibitemOpen
  \bibfield  {author} {\bibinfo {author} {\bibfnamefont {P.}~\bibnamefont
  {Rickhaus}}, \bibinfo {author} {\bibfnamefont {P.}~\bibnamefont {Makk}},
  \bibinfo {author} {\bibfnamefont {M.-H.}\ \bibnamefont {Liu}}, \bibinfo
  {author} {\bibfnamefont {E.}~\bibnamefont {Tóvári}}, \bibinfo {author}
  {\bibfnamefont {M.}~\bibnamefont {Weiss}}, \bibinfo {author} {\bibfnamefont
  {R.}~\bibnamefont {Maurand}}, \bibinfo {author} {\bibfnamefont
  {K.}~\bibnamefont {Richter}},\ and\ \bibinfo {author} {\bibfnamefont
  {C.}~\bibnamefont {Schoenenberger}},\ }\href
  {https://doi.org/10.1038/ncomms7470} {\bibfield  {journal} {\bibinfo
  {journal} {Nature Communications}\ }\textbf {\bibinfo {volume} {6}},\
  \bibinfo {pages} {6470} (\bibinfo {year} {2015})}\BibitemShut {NoStop}%
\bibitem [{\citenamefont {Chen}\ \emph {et~al.}(2016)\citenamefont {Chen},
  \citenamefont {Han}, \citenamefont {Elahi}, \citenamefont {Habib},
  \citenamefont {Wang}, \citenamefont {Wen}, \citenamefont {Gao}, \citenamefont
  {Taniguchi}, \citenamefont {Watanabe}, \citenamefont {Hone}, \citenamefont
  {Ghosh},\ and\ \citenamefont {Dean}}]{chen2016}%
  \BibitemOpen
  \bibfield  {author} {\bibinfo {author} {\bibfnamefont {S.}~\bibnamefont
  {Chen}}, \bibinfo {author} {\bibfnamefont {Z.}~\bibnamefont {Han}}, \bibinfo
  {author} {\bibfnamefont {M.~M.}\ \bibnamefont {Elahi}}, \bibinfo {author}
  {\bibfnamefont {K.~M.~M.}\ \bibnamefont {Habib}}, \bibinfo {author}
  {\bibfnamefont {L.}~\bibnamefont {Wang}}, \bibinfo {author} {\bibfnamefont
  {B.}~\bibnamefont {Wen}}, \bibinfo {author} {\bibfnamefont {Y.}~\bibnamefont
  {Gao}}, \bibinfo {author} {\bibfnamefont {T.}~\bibnamefont {Taniguchi}},
  \bibinfo {author} {\bibfnamefont {K.}~\bibnamefont {Watanabe}}, \bibinfo
  {author} {\bibfnamefont {J.}~\bibnamefont {Hone}}, \bibinfo {author}
  {\bibfnamefont {A.~W.}\ \bibnamefont {Ghosh}},\ and\ \bibinfo {author}
  {\bibfnamefont {C.~R.}\ \bibnamefont {Dean}},\ }\href
  {https://doi.org/10.1126/science.aaf5481} {\bibfield  {journal} {\bibinfo
  {journal} {Science}\ }\textbf {\bibinfo {volume} {353}},\ \bibinfo {pages}
  {1522} (\bibinfo {year} {2016})}\BibitemShut {NoStop}%
\bibitem [{\citenamefont {Kolasi\ifmmode~\acute{n}\else \'{n}\fi{}ski}\ \emph
  {et~al.}(2017)\citenamefont {Kolasi\ifmmode~\acute{n}\else \'{n}\fi{}ski},
  \citenamefont {Mre\ifmmode \acute{n}\else
  \'{n}\fi{}ca-Kolasi\ifmmode~\acute{n}\else \'{n}\fi{}ska},\ and\
  \citenamefont {Szafran}}]{kolasinski2017}%
  \BibitemOpen
  \bibfield  {author} {\bibinfo {author} {\bibfnamefont {K.}~\bibnamefont
  {Kolasi\ifmmode~\acute{n}\else \'{n}\fi{}ski}}, \bibinfo {author}
  {\bibfnamefont {A.}~\bibnamefont {Mre\ifmmode \acute{n}\else
  \'{n}\fi{}ca-Kolasi\ifmmode~\acute{n}\else \'{n}\fi{}ska}},\ and\ \bibinfo
  {author} {\bibfnamefont {B.}~\bibnamefont {Szafran}},\ }\href
  {https://doi.org/10.1103/PhysRevB.95.045304} {\bibfield  {journal} {\bibinfo
  {journal} {Phys. Rev. B}\ }\textbf {\bibinfo {volume} {95}},\ \bibinfo
  {pages} {045304} (\bibinfo {year} {2017})}\BibitemShut {NoStop}%
\bibitem [{\citenamefont {Makk}\ \emph {et~al.}(2018)\citenamefont {Makk},
  \citenamefont {Handschin}, \citenamefont {T\'ov\'ari}, \citenamefont
  {Watanabe}, \citenamefont {Taniguchi}, \citenamefont {Richter}, \citenamefont
  {Liu},\ and\ \citenamefont {Sch\"onenberger}}]{makk2018}%
  \BibitemOpen
  \bibfield  {author} {\bibinfo {author} {\bibfnamefont {P.}~\bibnamefont
  {Makk}}, \bibinfo {author} {\bibfnamefont {C.}~\bibnamefont {Handschin}},
  \bibinfo {author} {\bibfnamefont {E.}~\bibnamefont {T\'ov\'ari}}, \bibinfo
  {author} {\bibfnamefont {K.}~\bibnamefont {Watanabe}}, \bibinfo {author}
  {\bibfnamefont {T.}~\bibnamefont {Taniguchi}}, \bibinfo {author}
  {\bibfnamefont {K.}~\bibnamefont {Richter}}, \bibinfo {author} {\bibfnamefont
  {M.-H.}\ \bibnamefont {Liu}},\ and\ \bibinfo {author} {\bibfnamefont
  {C.}~\bibnamefont {Sch\"onenberger}},\ }\href
  {https://doi.org/10.1103/PhysRevB.98.035413} {\bibfield  {journal} {\bibinfo
  {journal} {Phys. Rev. B}\ }\textbf {\bibinfo {volume} {98}},\ \bibinfo
  {pages} {035413} (\bibinfo {year} {2018})}\BibitemShut {NoStop}%
\bibitem [{\citenamefont {Barbier}\ \emph {et~al.}(2012)\citenamefont
  {Barbier}, \citenamefont {Papp},\ and\ \citenamefont
  {Peeters}}]{barbier2012}%
  \BibitemOpen
  \bibfield  {author} {\bibinfo {author} {\bibfnamefont {M.}~\bibnamefont
  {Barbier}}, \bibinfo {author} {\bibfnamefont {G.}~\bibnamefont {Papp}},\ and\
  \bibinfo {author} {\bibfnamefont {F.~M.}\ \bibnamefont {Peeters}},\ }\href
  {https://doi.org/10.1063/1.4704667} {\bibfield  {journal} {\bibinfo
  {journal} {Applied Physics Letters}\ }\textbf {\bibinfo {volume} {100}},\
  \bibinfo {pages} {163121} (\bibinfo {year} {2012})}\BibitemShut {NoStop}%
\bibitem [{\citenamefont {Cheianov}\ and\ \citenamefont
  {Fal'ko}(2006)}]{cheianov2006}%
  \BibitemOpen
  \bibfield  {author} {\bibinfo {author} {\bibfnamefont {V.~V.}\ \bibnamefont
  {Cheianov}}\ and\ \bibinfo {author} {\bibfnamefont {V.~I.}\ \bibnamefont
  {Fal'ko}},\ }\href {https://doi.org/10.1103/PhysRevB.74.041403} {\bibfield
  {journal} {\bibinfo  {journal} {Phys. Rev. B}\ }\textbf {\bibinfo {volume}
  {74}},\ \bibinfo {pages} {041403} (\bibinfo {year} {2006})}\BibitemShut
  {NoStop}%
\bibitem [{\citenamefont {Young}\ and\ \citenamefont {Kim}(2009)}]{young2009}%
  \BibitemOpen
  \bibfield  {author} {\bibinfo {author} {\bibfnamefont {A.~F.}\ \bibnamefont
  {Young}}\ and\ \bibinfo {author} {\bibfnamefont {P.}~\bibnamefont {Kim}},\
  }\href@noop {} {\bibfield  {journal} {\bibinfo  {journal} {Nature Physics}\
  }\textbf {\bibinfo {volume} {5}},\ \bibinfo {pages} {222} (\bibinfo {year}
  {2009})}\BibitemShut {NoStop}%
\bibitem [{\citenamefont {Tudorovskiy}\ \emph {et~al.}(2012)\citenamefont
  {Tudorovskiy}, \citenamefont {Reijnders},\ and\ \citenamefont
  {Katsnelson}}]{tudorovskiy2012}%
  \BibitemOpen
  \bibfield  {author} {\bibinfo {author} {\bibfnamefont {T.}~\bibnamefont
  {Tudorovskiy}}, \bibinfo {author} {\bibfnamefont {K.~J.~A.}\ \bibnamefont
  {Reijnders}},\ and\ \bibinfo {author} {\bibfnamefont {M.~I.}\ \bibnamefont
  {Katsnelson}},\ }\href@noop {} {\bibfield  {journal} {\bibinfo  {journal}
  {Physica Scripta}\ }\textbf {\bibinfo {volume} {T146}},\ \bibinfo {pages}
  {014010} (\bibinfo {year} {2012})}\BibitemShut {NoStop}%
\bibitem [{\citenamefont {Abanin}\ \emph {et~al.}(2007)\citenamefont {Abanin},
  \citenamefont {Novoselov}, \citenamefont {Zeitler}, \citenamefont {Lee},
  \citenamefont {Geim},\ and\ \citenamefont {Levitov}}]{abanin2007}%
  \BibitemOpen
  \bibfield  {author} {\bibinfo {author} {\bibfnamefont {D.~A.}\ \bibnamefont
  {Abanin}}, \bibinfo {author} {\bibfnamefont {K.~S.}\ \bibnamefont
  {Novoselov}}, \bibinfo {author} {\bibfnamefont {U.}~\bibnamefont {Zeitler}},
  \bibinfo {author} {\bibfnamefont {P.~A.}\ \bibnamefont {Lee}}, \bibinfo
  {author} {\bibfnamefont {A.~K.}\ \bibnamefont {Geim}},\ and\ \bibinfo
  {author} {\bibfnamefont {L.~S.}\ \bibnamefont {Levitov}},\ }\href
  {https://doi.org/10.1103/PhysRevLett.98.196806} {\bibfield  {journal}
  {\bibinfo  {journal} {Phys. Rev. Lett.}\ }\textbf {\bibinfo {volume} {98}},\
  \bibinfo {pages} {196806} (\bibinfo {year} {2007})}\BibitemShut {NoStop}%
\bibitem [{\citenamefont {Xiang}\ \emph {et~al.}(2016)\citenamefont {Xiang},
  \citenamefont {Mre\ifmmode \acute{n}\else
  \'{n}\fi{}ca-Kolasi\ifmmode~\acute{n}\else \'{n}\fi{}ska}, \citenamefont
  {Miseikis}, \citenamefont {Guiducci}, \citenamefont
  {Kolasi\ifmmode~\acute{n}\else \'{n}\fi{}ski}, \citenamefont {Coletti},
  \citenamefont {Szafran}, \citenamefont {Beltram}, \citenamefont {Roddaro},\
  and\ \citenamefont {Heun}}]{xiang2016}%
  \BibitemOpen
  \bibfield  {author} {\bibinfo {author} {\bibfnamefont {S.}~\bibnamefont
  {Xiang}}, \bibinfo {author} {\bibfnamefont {A.}~\bibnamefont {Mre\ifmmode
  \acute{n}\else \'{n}\fi{}ca-Kolasi\ifmmode~\acute{n}\else \'{n}\fi{}ska}},
  \bibinfo {author} {\bibfnamefont {V.}~\bibnamefont {Miseikis}}, \bibinfo
  {author} {\bibfnamefont {S.}~\bibnamefont {Guiducci}}, \bibinfo {author}
  {\bibfnamefont {K.}~\bibnamefont {Kolasi\ifmmode~\acute{n}\else
  \'{n}\fi{}ski}}, \bibinfo {author} {\bibfnamefont {C.}~\bibnamefont
  {Coletti}}, \bibinfo {author} {\bibfnamefont {B.}~\bibnamefont {Szafran}},
  \bibinfo {author} {\bibfnamefont {F.}~\bibnamefont {Beltram}}, \bibinfo
  {author} {\bibfnamefont {S.}~\bibnamefont {Roddaro}},\ and\ \bibinfo {author}
  {\bibfnamefont {S.}~\bibnamefont {Heun}},\ }\href
  {https://doi.org/10.1103/PhysRevB.94.155446} {\bibfield  {journal} {\bibinfo
  {journal} {Phys. Rev. B}\ }\textbf {\bibinfo {volume} {94}},\ \bibinfo
  {pages} {155446} (\bibinfo {year} {2016})}\BibitemShut {NoStop}%
\bibitem [{\citenamefont {Marguerite}\ \emph {et~al.}(2019)\citenamefont
  {Marguerite}, \citenamefont {Birkbeck}, \citenamefont {Aharon-Steinberg},
  \citenamefont {Halbertal}, \citenamefont {Bagani}, \citenamefont {Marcus},
  \citenamefont {Myasoedov}, \citenamefont {Geim}, \citenamefont {Perello},\
  and\ \citenamefont {Zeldov}}]{marguerite2019}%
  \BibitemOpen
  \bibfield  {author} {\bibinfo {author} {\bibfnamefont {A.}~\bibnamefont
  {Marguerite}}, \bibinfo {author} {\bibfnamefont {J.}~\bibnamefont
  {Birkbeck}}, \bibinfo {author} {\bibfnamefont {A.}~\bibnamefont
  {Aharon-Steinberg}}, \bibinfo {author} {\bibfnamefont {D.}~\bibnamefont
  {Halbertal}}, \bibinfo {author} {\bibfnamefont {K.}~\bibnamefont {Bagani}},
  \bibinfo {author} {\bibfnamefont {I.}~\bibnamefont {Marcus}}, \bibinfo
  {author} {\bibfnamefont {Y.}~\bibnamefont {Myasoedov}}, \bibinfo {author}
  {\bibfnamefont {A.~K.}\ \bibnamefont {Geim}}, \bibinfo {author}
  {\bibfnamefont {D.~J.}\ \bibnamefont {Perello}},\ and\ \bibinfo {author}
  {\bibfnamefont {E.}~\bibnamefont {Zeldov}},\ }\href@noop {} {\bibfield
  {journal} {\bibinfo  {journal} {Nature}\ }\textbf {\bibinfo {volume} {575}},\
  \bibinfo {pages} {628} (\bibinfo {year} {2019})}\BibitemShut {NoStop}%
\bibitem [{\citenamefont {Moreau}\ \emph {et~al.}(2021)\citenamefont {Moreau},
  \citenamefont {Brun}, \citenamefont {Somanchi}, \citenamefont {Watanabe},
  \citenamefont {Taniguchi}, \citenamefont {Stampfer},\ and\ \citenamefont
  {Hackens}}]{moreau2021}%
  \BibitemOpen
  \bibfield  {author} {\bibinfo {author} {\bibfnamefont {N.}~\bibnamefont
  {Moreau}}, \bibinfo {author} {\bibfnamefont {B.}~\bibnamefont {Brun}},
  \bibinfo {author} {\bibfnamefont {S.}~\bibnamefont {Somanchi}}, \bibinfo
  {author} {\bibfnamefont {K.}~\bibnamefont {Watanabe}}, \bibinfo {author}
  {\bibfnamefont {T.}~\bibnamefont {Taniguchi}}, \bibinfo {author}
  {\bibfnamefont {C.}~\bibnamefont {Stampfer}},\ and\ \bibinfo {author}
  {\bibfnamefont {B.}~\bibnamefont {Hackens}},\ }\href@noop {} {\bibfield
  {journal} {\bibinfo  {journal} {Nature Communications}\ }\textbf {\bibinfo
  {volume} {12}} (\bibinfo {year} {2021})}\BibitemShut {NoStop}%
\bibitem [{\citenamefont {Kumar}\ \emph {et~al.}(2022)\citenamefont {Kumar},
  \citenamefont {Srivastav}, \citenamefont {Sp{\aa}nslätt}, \citenamefont
  {Watanabe}, \citenamefont {Taniguchi}, \citenamefont {Gefen}, \citenamefont
  {Mirlin},\ and\ \citenamefont {Das}}]{kumar2022}%
  \BibitemOpen
  \bibfield  {author} {\bibinfo {author} {\bibfnamefont {R.}~\bibnamefont
  {Kumar}}, \bibinfo {author} {\bibfnamefont {S.~K.}\ \bibnamefont
  {Srivastav}}, \bibinfo {author} {\bibfnamefont {C.}~\bibnamefont
  {Sp{\aa}nslätt}}, \bibinfo {author} {\bibfnamefont {K.}~\bibnamefont
  {Watanabe}}, \bibinfo {author} {\bibfnamefont {T.}~\bibnamefont {Taniguchi}},
  \bibinfo {author} {\bibfnamefont {Y.}~\bibnamefont {Gefen}}, \bibinfo
  {author} {\bibfnamefont {A.~D.}\ \bibnamefont {Mirlin}},\ and\ \bibinfo
  {author} {\bibfnamefont {A.}~\bibnamefont {Das}},\ }\href@noop {} {\bibfield
  {journal} {\bibinfo  {journal} {Nature Communications}\ }\textbf {\bibinfo
  {volume} {13}} (\bibinfo {year} {2022})}\BibitemShut {NoStop}%
\bibitem [{\citenamefont {Schaibley}\ \emph {et~al.}(2016)\citenamefont
  {Schaibley}, \citenamefont {Yu}, \citenamefont {Clark}, \citenamefont
  {Rivera}, \citenamefont {Ross}, \citenamefont {Seyler}, \citenamefont {Yao},\
  and\ \citenamefont {Xu}}]{schaibley2016}%
  \BibitemOpen
  \bibfield  {author} {\bibinfo {author} {\bibfnamefont {J.}~\bibnamefont
  {Schaibley}}, \bibinfo {author} {\bibfnamefont {H.}~\bibnamefont {Yu}},
  \bibinfo {author} {\bibfnamefont {G.}~\bibnamefont {Clark}}, \bibinfo
  {author} {\bibfnamefont {P.}~\bibnamefont {Rivera}}, \bibinfo {author}
  {\bibfnamefont {J.}~\bibnamefont {Ross}}, \bibinfo {author} {\bibfnamefont
  {K.}~\bibnamefont {Seyler}}, \bibinfo {author} {\bibfnamefont
  {W.}~\bibnamefont {Yao}},\ and\ \bibinfo {author} {\bibfnamefont
  {X.}~\bibnamefont {Xu}},\ }\href {https://doi.org/10.1038/natrevmats.2016.55}
  {\bibfield  {journal} {\bibinfo  {journal} {Nature Reviews Materials}\
  }\textbf {\bibinfo {volume} {1}} (\bibinfo {year} {2016})}\BibitemShut
  {NoStop}%
\bibitem [{\citenamefont {Morikawa}\ \emph {et~al.}(2015)\citenamefont
  {Morikawa}, \citenamefont {Masubuchi}, \citenamefont {Moriya}, \citenamefont
  {Watanabe}, \citenamefont {Taniguchi},\ and\ \citenamefont
  {Machida}}]{morikawa2015}%
  \BibitemOpen
  \bibfield  {author} {\bibinfo {author} {\bibfnamefont {S.}~\bibnamefont
  {Morikawa}}, \bibinfo {author} {\bibfnamefont {S.}~\bibnamefont {Masubuchi}},
  \bibinfo {author} {\bibfnamefont {R.}~\bibnamefont {Moriya}}, \bibinfo
  {author} {\bibfnamefont {K.}~\bibnamefont {Watanabe}}, \bibinfo {author}
  {\bibfnamefont {T.}~\bibnamefont {Taniguchi}},\ and\ \bibinfo {author}
  {\bibfnamefont {T.}~\bibnamefont {Machida}},\ }\href
  {https://doi.org/10.1063/1.4919380} {\bibfield  {journal} {\bibinfo
  {journal} {Applied Physics Letters}\ }\textbf {\bibinfo {volume} {106}},\
  \bibinfo {pages} {183101} (\bibinfo {year} {2015})}\BibitemShut {NoStop}%
\bibitem [{\citenamefont {Trifunovic}\ and\ \citenamefont
  {Brouwer}(2019)}]{trifunovic2019}%
  \BibitemOpen
  \bibfield  {author} {\bibinfo {author} {\bibfnamefont {L.}~\bibnamefont
  {Trifunovic}}\ and\ \bibinfo {author} {\bibfnamefont {P.~W.}\ \bibnamefont
  {Brouwer}},\ }\href {https://doi.org/10.1103/PhysRevB.99.205431} {\bibfield
  {journal} {\bibinfo  {journal} {Phys. Rev. B}\ }\textbf {\bibinfo {volume}
  {99}},\ \bibinfo {pages} {205431} (\bibinfo {year} {2019})}\BibitemShut
  {NoStop}%
\bibitem [{\citenamefont {Flór}\ \emph {et~al.}(2022)\citenamefont {Flór},
  \citenamefont {Lacerda-Santos}, \citenamefont {Fleury}, \citenamefont
  {Roulleau},\ and\ \citenamefont {Waintal}}]{flor2022}%
  \BibitemOpen
  \bibfield  {author} {\bibinfo {author} {\bibfnamefont {I.~M.}\ \bibnamefont
  {Flór}}, \bibinfo {author} {\bibfnamefont {A.}~\bibnamefont
  {Lacerda-Santos}}, \bibinfo {author} {\bibfnamefont {G.}~\bibnamefont
  {Fleury}}, \bibinfo {author} {\bibfnamefont {P.}~\bibnamefont {Roulleau}},\
  and\ \bibinfo {author} {\bibfnamefont {X.}~\bibnamefont {Waintal}},\
  }\href@noop {} {} (\bibinfo {year} {2022})\BibitemShut {NoStop}%
\bibitem [{\citenamefont {Kotilahti}\ \emph {et~al.}(2021)\citenamefont
  {Kotilahti}, \citenamefont {Burset}, \citenamefont {Moskalets},\ and\
  \citenamefont {Flindt}}]{kotilahti2021}%
  \BibitemOpen
  \bibfield  {author} {\bibinfo {author} {\bibfnamefont {J.}~\bibnamefont
  {Kotilahti}}, \bibinfo {author} {\bibfnamefont {P.}~\bibnamefont {Burset}},
  \bibinfo {author} {\bibfnamefont {M.}~\bibnamefont {Moskalets}},\ and\
  \bibinfo {author} {\bibfnamefont {C.}~\bibnamefont {Flindt}},\ }\href@noop {}
  {\bibfield  {journal} {\bibinfo  {journal} {Entropy}\ }\textbf {\bibinfo
  {volume} {23}},\ \bibinfo {pages} {736} (\bibinfo {year} {2021})}\BibitemShut
  {NoStop}%
\bibitem [{\citenamefont {Kramer}\ \emph {et~al.}(2010)\citenamefont {Kramer},
  \citenamefont {Kreisbeck},\ and\ \citenamefont {Krueckl}}]{kramer2010}%
  \BibitemOpen
  \bibfield  {author} {\bibinfo {author} {\bibfnamefont {T.}~\bibnamefont
  {Kramer}}, \bibinfo {author} {\bibfnamefont {C.}~\bibnamefont {Kreisbeck}},\
  and\ \bibinfo {author} {\bibfnamefont {V.}~\bibnamefont {Krueckl}},\
  }\href@noop {} {\bibfield  {journal} {\bibinfo  {journal} {Physica Scripta}\
  }\textbf {\bibinfo {volume} {82}},\ \bibinfo {pages} {038101} (\bibinfo
  {year} {2010})}\BibitemShut {NoStop}%
\bibitem [{\citenamefont {Chaves}\ \emph {et~al.}(2015)\citenamefont {Chaves},
  \citenamefont {Farias}, \citenamefont {Peeters},\ and\ \citenamefont
  {Ferreira}}]{chaves2015}%
  \BibitemOpen
  \bibfield  {author} {\bibinfo {author} {\bibfnamefont {A.}~\bibnamefont
  {Chaves}}, \bibinfo {author} {\bibfnamefont {G.~A.}\ \bibnamefont {Farias}},
  \bibinfo {author} {\bibfnamefont {F.~M.}\ \bibnamefont {Peeters}},\ and\
  \bibinfo {author} {\bibfnamefont {R.}~\bibnamefont {Ferreira}},\ }\href@noop
  {} {\bibfield  {journal} {\bibinfo  {journal} {Communications in
  Computational Physics}\ }\textbf {\bibinfo {volume} {17}},\ \bibinfo {pages}
  {850–866} (\bibinfo {year} {2015})}\BibitemShut {NoStop}%
\bibitem [{\citenamefont {Grasselli}\ \emph {et~al.}(2015)\citenamefont
  {Grasselli}, \citenamefont {Bertoni},\ and\ \citenamefont
  {Goldoni}}]{grasselli2015}%
  \BibitemOpen
  \bibfield  {author} {\bibinfo {author} {\bibfnamefont {F.}~\bibnamefont
  {Grasselli}}, \bibinfo {author} {\bibfnamefont {A.}~\bibnamefont {Bertoni}},\
  and\ \bibinfo {author} {\bibfnamefont {G.}~\bibnamefont {Goldoni}},\
  }\href@noop {} {\bibfield  {journal} {\bibinfo  {journal} {Journal of
  Chemical Physics}\ }\textbf {\bibinfo {volume} {142}} (\bibinfo {year}
  {2015})}\BibitemShut {NoStop}%
\bibitem [{\citenamefont {Grasselli}\ \emph {et~al.}(2016)\citenamefont
  {Grasselli}, \citenamefont {Bertoni},\ and\ \citenamefont
  {Goldoni}}]{grasselli2016}%
  \BibitemOpen
  \bibfield  {author} {\bibinfo {author} {\bibfnamefont {F.}~\bibnamefont
  {Grasselli}}, \bibinfo {author} {\bibfnamefont {A.}~\bibnamefont {Bertoni}},\
  and\ \bibinfo {author} {\bibfnamefont {G.}~\bibnamefont {Goldoni}},\ }\href
  {https://doi.org/10.1103/PhysRevB.93.195310} {\bibfield  {journal} {\bibinfo
  {journal} {Phys. Rev. B}\ }\textbf {\bibinfo {volume} {93}},\ \bibinfo
  {pages} {195310} (\bibinfo {year} {2016})}\BibitemShut {NoStop}%
\bibitem [{\citenamefont {Beggi}\ \emph
  {et~al.}(2015{\natexlab{a}})\citenamefont {Beggi}, \citenamefont {Bordone},
  \citenamefont {Buscemi},\ and\ \citenamefont {Bertoni}}]{beggi2015}%
  \BibitemOpen
  \bibfield  {author} {\bibinfo {author} {\bibfnamefont {A.}~\bibnamefont
  {Beggi}}, \bibinfo {author} {\bibfnamefont {P.}~\bibnamefont {Bordone}},
  \bibinfo {author} {\bibfnamefont {F.}~\bibnamefont {Buscemi}},\ and\ \bibinfo
  {author} {\bibfnamefont {A.}~\bibnamefont {Bertoni}},\ }\href@noop {}
  {\bibfield  {journal} {\bibinfo  {journal} {Journal of Physics: Condensed
  Matter}\ }\textbf {\bibinfo {volume} {27}},\ \bibinfo {pages} {475301}
  (\bibinfo {year} {2015}{\natexlab{a}})}\BibitemShut {NoStop}%
\bibitem [{\citenamefont {Beggi}\ \emph
  {et~al.}(2015{\natexlab{b}})\citenamefont {Beggi}, \citenamefont {Bertoni},\
  and\ \citenamefont {Bordone}}]{beggi2015_2}%
  \BibitemOpen
  \bibfield  {author} {\bibinfo {author} {\bibfnamefont {A.}~\bibnamefont
  {Beggi}}, \bibinfo {author} {\bibfnamefont {A.}~\bibnamefont {Bertoni}},\
  and\ \bibinfo {author} {\bibfnamefont {P.}~\bibnamefont {Bordone}},\
  }\href@noop {} {\bibfield  {journal} {\bibinfo  {journal} {Journal of
  Physics: Conference Series}\ }\textbf {\bibinfo {volume} {647}},\ \bibinfo
  {pages} {012023} (\bibinfo {year} {2015}{\natexlab{b}})}\BibitemShut
  {NoStop}%
\bibitem [{\citenamefont {Bellentani}\ \emph {et~al.}(2018)\citenamefont
  {Bellentani}, \citenamefont {Beggi}, \citenamefont {Bordone},\ and\
  \citenamefont {Bertoni}}]{bellentani2018}%
  \BibitemOpen
  \bibfield  {author} {\bibinfo {author} {\bibfnamefont {L.}~\bibnamefont
  {Bellentani}}, \bibinfo {author} {\bibfnamefont {A.}~\bibnamefont {Beggi}},
  \bibinfo {author} {\bibfnamefont {P.}~\bibnamefont {Bordone}},\ and\ \bibinfo
  {author} {\bibfnamefont {A.}~\bibnamefont {Bertoni}},\ }\href
  {https://doi.org/10.1103/PhysRevB.97.205419} {\bibfield  {journal} {\bibinfo
  {journal} {Phys. Rev. B}\ }\textbf {\bibinfo {volume} {97}},\ \bibinfo
  {pages} {205419} (\bibinfo {year} {2018})}\BibitemShut {NoStop}%
\bibitem [{\citenamefont {Bordone}\ \emph {et~al.}(2019)\citenamefont
  {Bordone}, \citenamefont {Bellentani},\ and\ \citenamefont
  {Bertoni}}]{bordone2019}%
  \BibitemOpen
  \bibfield  {author} {\bibinfo {author} {\bibfnamefont {P.}~\bibnamefont
  {Bordone}}, \bibinfo {author} {\bibfnamefont {L.}~\bibnamefont
  {Bellentani}},\ and\ \bibinfo {author} {\bibfnamefont {A.}~\bibnamefont
  {Bertoni}},\ }\href@noop {} {\bibfield  {journal} {\bibinfo  {journal}
  {Semiconductor Science and Technology}\ }\textbf {\bibinfo {volume} {34}},\
  \bibinfo {pages} {103001} (\bibinfo {year} {2019})}\BibitemShut {NoStop}%
\bibitem [{\citenamefont {Bellentani}\ \emph {et~al.}(2020)\citenamefont
  {Bellentani}, \citenamefont {Forghieri}, \citenamefont {Bordone},\ and\
  \citenamefont {Bertoni}}]{bellentani2020}%
  \BibitemOpen
  \bibfield  {author} {\bibinfo {author} {\bibfnamefont {L.}~\bibnamefont
  {Bellentani}}, \bibinfo {author} {\bibfnamefont {G.}~\bibnamefont
  {Forghieri}}, \bibinfo {author} {\bibfnamefont {P.}~\bibnamefont {Bordone}},\
  and\ \bibinfo {author} {\bibfnamefont {A.}~\bibnamefont {Bertoni}},\
  }\href@noop {} {\bibfield  {journal} {\bibinfo  {journal} {Physical Review
  B}\ }\textbf {\bibinfo {volume} {102}} (\bibinfo {year} {2020})}\BibitemShut
  {NoStop}%
\bibitem [{\citenamefont {Hancock}\ \emph {et~al.}(2010)\citenamefont
  {Hancock}, \citenamefont {Uppstu}, \citenamefont {Saloriutta}, \citenamefont
  {Harju},\ and\ \citenamefont {Puska}}]{hancock2010}%
  \BibitemOpen
  \bibfield  {author} {\bibinfo {author} {\bibfnamefont {Y.}~\bibnamefont
  {Hancock}}, \bibinfo {author} {\bibfnamefont {A.}~\bibnamefont {Uppstu}},
  \bibinfo {author} {\bibfnamefont {K.}~\bibnamefont {Saloriutta}}, \bibinfo
  {author} {\bibfnamefont {A.}~\bibnamefont {Harju}},\ and\ \bibinfo {author}
  {\bibfnamefont {M.~J.}\ \bibnamefont {Puska}},\ }\href
  {https://doi.org/10.1103/PhysRevB.81.245402} {\bibfield  {journal} {\bibinfo
  {journal} {Phys. Rev. B}\ }\textbf {\bibinfo {volume} {81}},\ \bibinfo
  {pages} {245402} (\bibinfo {year} {2010})}\BibitemShut {NoStop}%
\bibitem [{\citenamefont {Katsnelson}\ and\ \citenamefont
  {Novoselov}(2007)}]{katsnelson2007}%
  \BibitemOpen
  \bibfield  {author} {\bibinfo {author} {\bibfnamefont {M.}~\bibnamefont
  {Katsnelson}}\ and\ \bibinfo {author} {\bibfnamefont {K.}~\bibnamefont
  {Novoselov}},\ }\href
  {https://doi.org/https://doi.org/10.1016/j.ssc.2007.02.043} {\bibfield
  {journal} {\bibinfo  {journal} {Solid State Communications}\ }\textbf
  {\bibinfo {volume} {143}},\ \bibinfo {pages} {3} (\bibinfo {year}
  {2007})}\BibitemShut {NoStop}%
\bibitem [{\citenamefont {Goerbig}(2011)}]{goerbig2011}%
  \BibitemOpen
  \bibfield  {author} {\bibinfo {author} {\bibfnamefont {M.~O.}\ \bibnamefont
  {Goerbig}},\ }\href {https://doi.org/10.1103/RevModPhys.83.1193} {\bibfield
  {journal} {\bibinfo  {journal} {Rev. Mod. Phys.}\ }\textbf {\bibinfo {volume}
  {83}},\ \bibinfo {pages} {1193} (\bibinfo {year} {2011})}\BibitemShut
  {NoStop}%
\bibitem [{\citenamefont {Lado}\ \emph {et~al.}(2015)\citenamefont {Lado},
  \citenamefont {García-Martínez},\ and\ \citenamefont
  {Fernández-Rossier}}]{lado2015}%
  \BibitemOpen
  \bibfield  {author} {\bibinfo {author} {\bibfnamefont {J.}~\bibnamefont
  {Lado}}, \bibinfo {author} {\bibfnamefont {N.}~\bibnamefont
  {García-Martínez}},\ and\ \bibinfo {author} {\bibfnamefont
  {J.}~\bibnamefont {Fernández-Rossier}},\ }\href
  {https://doi.org/https://doi.org/10.1016/j.synthmet.2015.06.026} {\bibfield
  {journal} {\bibinfo  {journal} {Synthetic Metals}\ }\textbf {\bibinfo
  {volume} {210}},\ \bibinfo {pages} {56} (\bibinfo {year} {2015})}\BibitemShut
  {NoStop}%
\bibitem [{\citenamefont {Marconcini}\ \emph {et~al.}(2010)\citenamefont
  {Marconcini}, \citenamefont {Logoteta}, \citenamefont {Fagotti},\ and\
  \citenamefont {Macucci}}]{marconcini2010}%
  \BibitemOpen
  \bibfield  {author} {\bibinfo {author} {\bibfnamefont {P.}~\bibnamefont
  {Marconcini}}, \bibinfo {author} {\bibfnamefont {D.}~\bibnamefont
  {Logoteta}}, \bibinfo {author} {\bibfnamefont {M.}~\bibnamefont {Fagotti}},\
  and\ \bibinfo {author} {\bibfnamefont {M.}~\bibnamefont {Macucci}},\ }in\
  \href {https://doi.org/10.1109/IWCE.2010.5677938} {\emph {\bibinfo
  {booktitle} {2010 14th International Workshop on Computational
  Electronics}}}\ (\bibinfo {year} {2010})\ pp.\ \bibinfo {pages}
  {1--4}\BibitemShut {NoStop}%
\bibitem [{\citenamefont {Brey}\ and\ \citenamefont
  {Fertig}(2006{\natexlab{a}})}]{brey2006}%
  \BibitemOpen
  \bibfield  {author} {\bibinfo {author} {\bibfnamefont {L.}~\bibnamefont
  {Brey}}\ and\ \bibinfo {author} {\bibfnamefont {H.~A.}\ \bibnamefont
  {Fertig}},\ }\href {https://doi.org/10.1103/PhysRevB.73.195408} {\bibfield
  {journal} {\bibinfo  {journal} {Phys. Rev. B}\ }\textbf {\bibinfo {volume}
  {73}},\ \bibinfo {pages} {195408} (\bibinfo {year}
  {2006}{\natexlab{a}})}\BibitemShut {NoStop}%
\bibitem [{\citenamefont {Brey}\ and\ \citenamefont
  {Fertig}(2006{\natexlab{b}})}]{brey2006_1}%
  \BibitemOpen
  \bibfield  {author} {\bibinfo {author} {\bibfnamefont {L.}~\bibnamefont
  {Brey}}\ and\ \bibinfo {author} {\bibfnamefont {H.~A.}\ \bibnamefont
  {Fertig}},\ }\href {https://doi.org/10.1103/PhysRevB.73.235411} {\bibfield
  {journal} {\bibinfo  {journal} {Phys. Rev. B}\ }\textbf {\bibinfo {volume}
  {73}},\ \bibinfo {pages} {235411} (\bibinfo {year}
  {2006}{\natexlab{b}})}\BibitemShut {NoStop}%
\bibitem [{sup()}]{supplemental}%
  \BibitemOpen
  \href@noop {} {\bibinfo {title} {See supplemental material for animations of
  the simulations presented in the text.}}\BibitemShut {Stop}%
\bibitem [{\citenamefont {Katsnelson}\ \emph {et~al.}(2006)\citenamefont
  {Katsnelson}, \citenamefont {Novoselov},\ and\ \citenamefont
  {Geim}}]{katsnelson2006}%
  \BibitemOpen
  \bibfield  {author} {\bibinfo {author} {\bibfnamefont {M.~I.}\ \bibnamefont
  {Katsnelson}}, \bibinfo {author} {\bibfnamefont {K.~S.}\ \bibnamefont
  {Novoselov}},\ and\ \bibinfo {author} {\bibfnamefont {A.~K.}\ \bibnamefont
  {Geim}},\ }\href@noop {} {\bibfield  {journal} {\bibinfo  {journal} {Nature
  Physics}\ }\textbf {\bibinfo {volume} {2}},\ \bibinfo {pages} {620} (\bibinfo
  {year} {2006})}\BibitemShut {NoStop}%
\bibitem [{\citenamefont {Ronen}\ \emph {et~al.}(2021)\citenamefont {Ronen},
  \citenamefont {Werkmeister}, \citenamefont {Haie~Najafabadi}, \citenamefont
  {Pierce}, \citenamefont {Anderson}, \citenamefont {Shin}, \citenamefont
  {Lee}, \citenamefont {Lee}, \citenamefont {Johnson}, \citenamefont
  {Watanabe}, \citenamefont {Taniguchi}, \citenamefont {Yacoby},\ and\
  \citenamefont {Kim}}]{ronen2021}%
  \BibitemOpen
  \bibfield  {author} {\bibinfo {author} {\bibfnamefont {Y.}~\bibnamefont
  {Ronen}}, \bibinfo {author} {\bibfnamefont {T.}~\bibnamefont {Werkmeister}},
  \bibinfo {author} {\bibfnamefont {D.}~\bibnamefont {Haie~Najafabadi}},
  \bibinfo {author} {\bibfnamefont {A.~T.}\ \bibnamefont {Pierce}}, \bibinfo
  {author} {\bibfnamefont {L.~E.}\ \bibnamefont {Anderson}}, \bibinfo {author}
  {\bibfnamefont {Y.~J.}\ \bibnamefont {Shin}}, \bibinfo {author}
  {\bibfnamefont {S.~Y.}\ \bibnamefont {Lee}}, \bibinfo {author} {\bibfnamefont
  {Y.~H.}\ \bibnamefont {Lee}}, \bibinfo {author} {\bibfnamefont
  {B.}~\bibnamefont {Johnson}}, \bibinfo {author} {\bibfnamefont
  {K.}~\bibnamefont {Watanabe}}, \bibinfo {author} {\bibfnamefont
  {T.}~\bibnamefont {Taniguchi}}, \bibinfo {author} {\bibfnamefont
  {A.}~\bibnamefont {Yacoby}},\ and\ \bibinfo {author} {\bibfnamefont
  {P.}~\bibnamefont {Kim}},\ }\href@noop {} {\bibfield  {journal} {\bibinfo
  {journal} {Nature Nanotechnology}\ }\textbf {\bibinfo {volume} {16}}
  (\bibinfo {year} {2021})}\BibitemShut {NoStop}%
\bibitem [{\citenamefont {Zeng}\ \emph {et~al.}(2019)\citenamefont {Zeng},
  \citenamefont {Li}, \citenamefont {Dietrich}, \citenamefont {Ghosh},
  \citenamefont {Watanabe}, \citenamefont {Taniguchi}, \citenamefont {Hone},\
  and\ \citenamefont {Dean}}]{zeng2019}%
  \BibitemOpen
  \bibfield  {author} {\bibinfo {author} {\bibfnamefont {Y.}~\bibnamefont
  {Zeng}}, \bibinfo {author} {\bibfnamefont {J.~I.~A.}\ \bibnamefont {Li}},
  \bibinfo {author} {\bibfnamefont {S.~A.}\ \bibnamefont {Dietrich}}, \bibinfo
  {author} {\bibfnamefont {O.~M.}\ \bibnamefont {Ghosh}}, \bibinfo {author}
  {\bibfnamefont {K.}~\bibnamefont {Watanabe}}, \bibinfo {author}
  {\bibfnamefont {T.}~\bibnamefont {Taniguchi}}, \bibinfo {author}
  {\bibfnamefont {J.}~\bibnamefont {Hone}},\ and\ \bibinfo {author}
  {\bibfnamefont {C.~R.}\ \bibnamefont {Dean}},\ }\href
  {https://doi.org/10.1103/PhysRevLett.122.137701} {\bibfield  {journal}
  {\bibinfo  {journal} {Phys. Rev. Lett.}\ }\textbf {\bibinfo {volume} {122}},\
  \bibinfo {pages} {137701} (\bibinfo {year} {2019})}\BibitemShut {NoStop}%
\bibitem [{\citenamefont {Polshyn}\ \emph {et~al.}(2018)\citenamefont
  {Polshyn}, \citenamefont {Zhou}, \citenamefont {Spanton}, \citenamefont
  {Taniguchi}, \citenamefont {Watanabe},\ and\ \citenamefont
  {Young}}]{polshyn2018}%
  \BibitemOpen
  \bibfield  {author} {\bibinfo {author} {\bibfnamefont {H.}~\bibnamefont
  {Polshyn}}, \bibinfo {author} {\bibfnamefont {H.}~\bibnamefont {Zhou}},
  \bibinfo {author} {\bibfnamefont {E.~M.}\ \bibnamefont {Spanton}}, \bibinfo
  {author} {\bibfnamefont {T.}~\bibnamefont {Taniguchi}}, \bibinfo {author}
  {\bibfnamefont {K.}~\bibnamefont {Watanabe}},\ and\ \bibinfo {author}
  {\bibfnamefont {A.~F.}\ \bibnamefont {Young}},\ }\href
  {https://doi.org/10.1103/PhysRevLett.121.226801} {\bibfield  {journal}
  {\bibinfo  {journal} {Phys. Rev. Lett.}\ }\textbf {\bibinfo {volume} {121}},\
  \bibinfo {pages} {226801} (\bibinfo {year} {2018})}\BibitemShut {NoStop}%
\bibitem [{\citenamefont {Chung}\ \emph {et~al.}(2005)\citenamefont {Chung},
  \citenamefont {Samuelsson},\ and\ \citenamefont {B\"uttiker}}]{chung2005}%
  \BibitemOpen
  \bibfield  {author} {\bibinfo {author} {\bibfnamefont {V.~S.-W.}\
  \bibnamefont {Chung}}, \bibinfo {author} {\bibfnamefont {P.}~\bibnamefont
  {Samuelsson}},\ and\ \bibinfo {author} {\bibfnamefont {M.}~\bibnamefont
  {B\"uttiker}},\ }\href {https://doi.org/10.1103/PhysRevB.72.125320}
  {\bibfield  {journal} {\bibinfo  {journal} {Phys. Rev. B}\ }\textbf {\bibinfo
  {volume} {72}},\ \bibinfo {pages} {125320} (\bibinfo {year}
  {2005})}\BibitemShut {NoStop}%
\bibitem [{\citenamefont {Giovannetti}\ \emph {et~al.}(2008)\citenamefont
  {Giovannetti}, \citenamefont {Taddei}, \citenamefont {Frustaglia},\ and\
  \citenamefont {Fazio}}]{giovannetti2008}%
  \BibitemOpen
  \bibfield  {author} {\bibinfo {author} {\bibfnamefont {V.}~\bibnamefont
  {Giovannetti}}, \bibinfo {author} {\bibfnamefont {F.}~\bibnamefont {Taddei}},
  \bibinfo {author} {\bibfnamefont {D.}~\bibnamefont {Frustaglia}},\ and\
  \bibinfo {author} {\bibfnamefont {R.}~\bibnamefont {Fazio}},\ }\href
  {https://doi.org/10.1103/PhysRevB.77.155320} {\bibfield  {journal} {\bibinfo
  {journal} {Phys. Rev. B}\ }\textbf {\bibinfo {volume} {77}},\ \bibinfo
  {pages} {155320} (\bibinfo {year} {2008})}\BibitemShut {NoStop}%
\end{thebibliography}%

\end{document}